# A comprehensive study of noble gases and nitrogen in "Hypatia", a diamond-rich pebble from SW Egypt

Guillaume Avice[a*], Matthias M. M. Meier[b, a], Bernard Marty[a], Rainer Wieler[b], Jan D. Kramers[c], Falko Langenhorst[d], Pierre Cartigny[e], Colin Maden[b], Laurent Zimmermann[a], Marco A. G. Andreoli[f]

***corresponding author. E-mail: gavice@crpg.cnrs-nancy.fr; Tel: +33 3 83 59 42 46**

[a]CRPG-CNRS, Université de Lorraine, UMR 7358, 15 rue Notre-Dame des Pauvres, BP 20, 54501 Vandoeuvre-lès-Nancy Cedex, France.

[b]Department of Earth Sciences, ETH Zürich, Clausiusstrasse 25, CH-8092 Zürich, Switzerland.

[c]Department of Geology, University of Johannesburg, Auckland Park 2006, Johannesburg, South Africa.

[d]Institut für Geowissenschaften, Friedrich-Schiller-Universität Jena, Carl-Zeiss-Promenade 10, D-07745 Jena, Germany.

[e]Équipe de géochimie des isotopes stables, Institut de Physique du Globe de Paris, Sorbonne Paris Cité, Univ. Paris Diderot, UMR 7154 CNRS, F-75005 Paris, France.

[f]School of Geosciences, University of the Witwatersrand, PO Box 3, Wits 2050, South Africa.

## Abstract

This is a follow-up study of a work by Kramers et al. (2013) on a very unusual diamond-rich rock fragment found in the area of south west Egypt in the south-western side of the Libyan Desert Glass strewn field. This pebble, called Hypatia, is composed of almost pure carbon. Transmission Electron Microscopy (TEM) and X-ray diffraction (XRD) results reveal that Hypatia is mainly made of defect-rich diamond containing lonsdaleite and multiple deformation bands. These characteristics are compatible with an impact origin on Earth and/or in space. We also analyzed concentrations and isotopic compositions of all five noble gases and nitrogen in several ~mg sized Hypatia samples. These data confirm the conclusion by Kramers et al. (2013) that Hypatia is extra-terrestrial. The sample is relatively rich in trapped noble gases with an isotopic composition being close to the Q component found in many types of meteorites. $^{40}Ar/^{36}Ar$ ratios in individual steps are as low as 0.4 ± 0.3. Cosmic-ray produced "cosmogenic" $^{21}Ne$ is present in concentrations corresponding to a nominal cosmic-ray exposure (CRE) age of roughly 0.1 Myr if produced in a typical meter-sized meteoroid. Such an atypically low nominal CRE age suggests high shielding in a considerably larger body. In addition to the Xe-Q composition, an excess of radiogenic $^{129}Xe$ (from the decay of short-lived radioactive $^{129}I$) is observed ($^{129}Xe/^{132}Xe$ = 1.18 +/- 0.03). Two isotopically distinct N components are present, an isotopically heavy component ($\delta^{15}N$ ~ +20‰) released at low temperatures and a major isotopically light component ($\delta^{15}N$ ~ -110‰) at higher temperatures. This disequilibrium in N suggests that the diamonds in Hypatia were formed in space rather than upon impact on Earth ($\delta^{15}N_{atm}$ = 0 ‰). All our data are broadly consistent with concentrations and isotopic compositions of noble gases in at least three different types of carbon-rich meteoritic materials: carbon-rich veins in ureilites, graphite in acapulcoites/lodranites and graphite nodules in iron meteorites. However, Hypatia does not seem to be directly related to any of these materials, but may have sampled a similar cosmochemical reservoir. Our study does not confirm the presence of exotic noble gases (e.g. G component) that led Kramers et al. (2013) to propose that Hypatia is a remnant of a comet nucleus that impacted the Earth.

## 1. Introduction

In 1996 a very unusual ~30 g sized pebble was found in the Libyan Desert Glass strewn field where abundant fragments of impact-related silica-rich glass are found (Barakat, 2012; Reimold and Koeberl, 2014). This brittle black stone (Fig. S1) consists of ~70 wt. % carbon, and has a hardness comparable to diamond, reminiscent of carbonados. Kramers et al. (2013) named the stone "Hypatia" in honor of a 4$^{th}$ century female philosopher from Alexandria (Egypt). These authors performed an exploratory analytical study on Hypatia, including XRD, SEM, Raman spectroscopy, TEM, and analyses of C and noble gas isotopes motivated by the fact that this stone was found in the area of the Libyan Desert Glass (LDG), the origin of which remains enigmatic (Reimold and Koeberl, 2014). Noble gas isotope analysis is central to the study of meteorites because these rocks formed from multiple components with distinct noble gas isotopic signatures that help to constrain their origin and evolution. Among these components, the so-called Q phase dominates the budget of heavy noble gases (Ar, Kr Xe) in chondrites originating from the asteroid belt. Although the chemical nature (Marrocchi et al., 2015) and mode of formation (Kuga et al., 2015; Ott, 2014) of Q are debated, this component is chemically and isotopically fractionated relative to the isotopic composition of the Solar Wind (Meshik et al., 2014) possibly due to ionization processes (Marrocchi et al., 2011) and is ubiquitous in pristine to moderately metamorphosed chondrites (Busemann et al., 2000). Other minor noble gas components (*e.g.* the P3 and G components) are found in presolar materials (for example SiC) trapped in meteorites (Ott, 2014). These components are derived from sources external to the solar system and carry isotopic signatures characteristic of nucleosynthesis in stars. Kramers et al. (2013) concluded that Hypatia is extra-terrestrial, based on $^{40}Ar/^{36}Ar$ ratios as low as about 40. They noted that O/C ratios (0.19 - 0.51) in Hypatia are higher than in chondritic Insoluble Organic Matter (IOM). In addition, they reported that the trapped Ne, Kr, and Xe in Hypatia indicate the occurrence of the nucleosynthetic P3 and G components of presolar origin known from meteorites (Ott, 2014), while the Q (and HL) components ubiquitous in chondrites were absent in Hypatia. The combined evidence led them to conclude that Hypatia did not originate in the asteroid belt where chondrites likely formed. They suggested instead that it formed in a more external region of the solar accretion disk, such as the Kuiper Belt, where presolar components might be more abundant, i.e., that Hypatia could be of cometary origin. They further proposed that the airburst of the parent comet of Hypatia resulted in the formation of the Libyan Desert Glass. This interpretation was subsequently criticized by Reimold and Koeberl (2014), although a cometary origin for the Libyan Desert Glass has been advocated many times, starting with Urey's seminal paper (Urey, 1957).

In this work we extend the study by Kramers et al. (2013) with isotopic analyses of all five noble gases in several mg-sized fragments of Hypatia in two different laboratories (CRPG Nancy, France and ETH Zürich, Switzerland) and with a nitrogen isotope investigation performed both at CRPG (Nancy) and IPG-Paris. We also describe results from X-ray diffraction (XRD) experiments and transmission electron microscopy (TEM) observations performed at the University of Jena (Germany). An attempt to determine the oxygen isotopic composition in Hypatia by the Nancy Cameca 1280 ion probe failed because of the reduced size of oxygen-bearing phases and because of the presence of contaminants and important amounts of water. This new study confirms and provides new evidence for the earlier conclusion that Hypatia is a fascinating new type of extra-terrestrial material. In contrast to the exploratory work reported by Kramers et al. (2013), we did find noble gases with isotopic signatures closely resembling the Q component. We also found nitrogen with an isotopic signature clearly distinct from primitive chondrites and closely resembling those of various differentiated meteoritic materials. In particular, we compare our data with noble gas and nitrogen signatures in three known types of carbon-rich extraterrestrial materials: carbon-rich veins in ureilite meteorites, graphite nodules in iron meteorites, and carbon-rich lithologies in acapulcoites and lodranites, and we discuss a possible link of Hypatia with each of these materials.

## 2. Samples and methods

X-ray diffraction and Transmission Electron Microscopy (TEM) techniques used in this study are described in the supplementary material.

### 2.1 Noble gas and nitrogen analyses

Table 1 indicates the samples analyzed and the analytical techniques used in Nancy, Zürich and Paris. Samples are from the same group of samples (≈1 g) as used by Kramers et al. (2013). In Nancy, noble gases and nitrogen were extracted conjointly in samples H-N1, H-N2 and H-N3 upon heating in high-vacuum with an infra-red (IR) $CO_2$ laser (Humbert et al., 2000). Extracted gases were subsequently split into an aliquot for nitrogen analysis and another one for noble gas analysis. The nitrogen was purified in a glass line and analyzed using a VG 5400 mass spectrometer following the procedure reported in (Zimmermann et al., 2009). The noble gas aliquot was purified using Ti-sponge getters and Ar was trapped on charcoal held at liquid $N_2$ temperature. Following N analysis, Ne and Ar were sequentially analyzed using the same mass spectrometer. Sample H-N4 was heated in five extraction steps (400, 850, 1400, 1800 and 2200 °C) in a Ta crucible using an induction furnace. Extracted gases were purified on three Ti-sponge getters. Xe and Kr were trapped on a quartz tube cooled to 77 K. Xe and Kr were then sequentially analyzed on a Helix MC Plus mass spectrometer. Mass spectrometer sensitivity was calibrated with known amounts of atmospheric noble gases following the same procedure as reported in (Marty and Zimmermann, 1999).

In Paris, the nitrogen content and isotope composition of two Hypatia diamond samples were investigated following methods described previously (Boyd et al., 1995). Importantly, the two analyzed Hypatia pieces (about 1.5 mg) were pre-combusted, over 6 hours at 600°C in order to remove any organic matter and/or graphite and, therefore, ensuring that the analyzed carbon phase consisted essentially of pure diamonds. The two samples were weighed before their analysis and $CO_2$ yield was used to determined the abundance of carbon.

In Zürich, noble gases were extracted from samples Z1 and Z2 with a 30 W continuous-wave IR laser ( $\lambda$ = 1064 nm) heating the samples for about 60 s (Vogel et al., 2003). Noble gases in samples Z3, Z4, and Z5 were extracted at ~1800 °C during 30 min in a Mo crucible heated by electron bombardment. For all analyses, the respective sample chamber had been preheated at 100 °C for ~24 hours to remove adsorbed atmospheric gases. Extracted gases were cleaned by various getters before Ar was frozen onto charcoal at liquid nitrogen temperature. The He-Ne and the Ar fractions were sequentially analyzed in a custom-built sector-field noble gas mass spectrometer equipped with an ion counting multiplier and a Faraday detector. Additional details of the analytical procedure are given in (Wieler et al., 1989). An additional step after the main extraction was done for samples H-Z1, H-Z2, and H-Z3 and demonstrated complete gas extraction during the main step. Blank gas amounts for the laser extractions (H-Z1 and H-Z2) were determined by firing the laser at an empty spot of the sample holder for 60 s. For the furnace extractions (for H-Z3 to H-Z5) a piece of Al foil of the same mass as used to wrap samples was melted at the same temperature and for the same duration as the samples. Mass spectrometer sensitivity was calibrated with known amounts of pure standard gases as described in (Heber et al., 2009).

## 3. Results

### 3.1 XRD results and TEM observations

The X-ray diffraction pattern of Hypatia is characterized by broad, low-intensity peaks (Figure S2) indicating the poor crystallinity of the material. A comparison with the pattern of well-crystallized synthetic diamond shows that the main X-ray reflections in the pattern of the Hypatia sample are fully compatible with the diamond structure. X-ray diffraction tails on both sides of the 111 diamond peak were also observed and can be interpreted as the $10\bar{1}0$ and $10\bar{1}1$ peaks of the hexagonal high-pressure polymorph lonsdaleite. The position of the 0002 peak of lonsdaleite would thereby coincide with the 111 peak of diamond. The lattice parameter *a* refined from the X-ray diffraction pattern of the Hypatia diamond is 3.57 Å, identical to the lattice constant of normal diamond. Besides diamond and lonsdaleite no other phase could be identified on the basis of X-ray diffraction.

X-ray line broadening analysis of Hypatia was additionally used to obtain clues to the causes for the poor crystallinity. This analysis revealed that the limited long-range order of Hypatia diamond is not only due to small crystallite size, but is also partly attributable to considerable internal strain, indicative of numerous lattice defects.

The defect microstructure was, therefore, examined by TEM. Conventional TEM imaging shows that Hypatia is composed of numerous diamond grains with sizes on the order of a few micrometers. Some grains exhibit multiple 100 nm wide bands with alternating diffraction contrast (Figure S3). Although this alternating contrast suggests a twin configuration, electron diffraction disproves one such possibility for the diamond structure. Similar multiple bands have only been found in impact diamonds from various craters and could be attributed to shock-induced mechanical twins of the graphite precursor (Langenhorst et al., 1999). The multiple twinning must have occurred just before the solid-state transformation when the shock wave entered into graphite (Langenhorst and Deutsch, 2012).

High-resolution TEM images reveal that diamond is associated with some graphite discernible by the typical 0001 spacing of 3.35 Å. The graphite can be onion-shaped and covers the surface of diamond (see Fig. S4 in supplementary material). The graphite-like areas are mostly < 10 nm in size explaining why they were not detected by X-ray diffraction.

The coating of diamond surfaces with graphite and the absence of an epitactic orientation relationship between the two phases (e.g. $(0001)_{graphite}$ does not coincide with$(111)_{diamond}$) suggest that the graphite flakes are probably a result of retrograde annealing of diamond.

### 3.2 Noble gas and nitrogen results

Concentrations and isotopic compositions of noble gases and nitrogen extracted in Nancy, Zürich and Paris are shown in Tables 2, 3 and S1 (supplementary material) respectively. Abundances and isotopic composition of nitrogen and carbon extracted in Paris are shown in supplementary material (Table S1). In this section we will first discuss isotopic compositions and then gas abundances and elemental ratios.

#### 3.2.1 Helium

The $^{3}He/^{4}He$ ratios in all five Hypatia "Z" samples are very similar to values observed for He-Q, with a weighted average of $(1.55 \pm 0.11) \times 10^{-4}$. While this value is slightly higher than the "canonical value" of He-Q of $\sim 1.23 \times 10^{-4}$ (Busemann et al., 2000; Ott, 2014) values around $\sim 1.6 \times 10^{-4}$ for He-Q-rich samples have also been reported (Wieler et al., 1991). Remarkably, the measured $^{3}He/^{4}He$ ratios do not vary systematically with $^{4}He$ concentration (the somewhat higher value of sample H-Z1 of $2.1 \times 10^{-4}$ has an untypically large uncertainty). This also suggests that He in Hypatia is essentially trapped and of Q origin, as otherwise contributions from cosmic-ray produced ($^{3}He$-rich) and radiogenic $^{4}He$ in each of the samples would have to add up fortuitously to a Q-like ratio. This seems very unlikely, although it needs to be noted that the $^{4}He/^{36}Ar$ ratios in Hypatia (~15-20) are higher than the mean value of ≈5 for Q (Ott, 2014). However, $^{4}He/^{36}Ar$ ratios in Q vary over a wide range of values (1-11; (Busemann et al., 2000)). To explain this latter observation by variable concentrations of radiogenic $^{4}He$ in different pieces of Hypatia would require U and Th concentrations of roughly half the values in ordinary chondrites (assuming an age of Hypatia of 4.56 Ga). This is unlikely given that only a few percent of Hypatia is non-carbonaceous material. Together with the fact that all analyzed samples show similar $^{3}He/^{4}He$ ratios, we can thus exclude that the Q-like value of He in Hypatia is due to radiogenic and cosmogenic He fortuitously adding up. The majority of the He in Hypatia must be trapped with an isotopic composition close to that of He-Q. This means that a reliable concentration of cosmogenic $^{3}He$ cannot be easily derived in any of our samples, as the exact isotopic composition of the trapped Q-like He remains unknown. See next section for a discussion about the abundance of cosmogenic $^{3}He$.

### 3.2.2 Neon

All temperature steps and total extractions are listed in Tables 2 and 3, and the totals and some selected data from individual steps are shown in Fig. 1. The $^{20}Ne/^{22}Ne$ ratios of the total gas measured in all three Nancy samples and the ratios of the three Zürich samples all vary between 10 and 11, with a weighted average of 11.0 ± 0.8. This average value is within the range of $^{20}Ne/^{22}Ne$ ratios of 10.1-10.7 reported for Ne-Q in many meteorite classes (Busemann et al., 2000; Ott, 2014). Although some individual temperature steps (Table 2, Fig. 1) likely indicate the release of some atmospheric Ne and one individual step in sample H-N2 gives a low value of 7.3±0.4 for $^{20}Ne/^{22}Ne$ (Table 2), we conclude that the major portion of the Ne in Hypatia represents Ne-Q. We find no clear evidence for the presence of Ne-HL or "exotic" Ne-G (essentially pure $^{22}Ne$) in Hypatia, as Kramers et al. (2013) reported.

A contribution of excess $^{21}Ne$ is visible in individual data points located at the right side of pure potential end-members (Fig. 1) and pointing towards the cosmogenic component ($^{21}Ne/^{22}Ne$=0.93, (Leya et al., 2001)). Concentrations of excess $^{21}Ne$ in the bulk individual samples range between $5x10^{-16}$ and $4x10^{-15}$ $mol.g^{-1}$. This excess $^{21}Ne$ cannot be nucleogenic, as can be estimated by conservatively assuming all measured $^{4}He$ to be radiogenic (see previous subsection 3.2.1) and a $^{21}Ne_{nuc}/^{4}He_{rad}$ ratio of ~2.8 x $10^{-8}$ for a O-content of ca. 30 % (Cox et al., 2015). Furthermore, the excess $^{21}Ne$ cannot have been produced on Earth by cosmic rays, as this would require an unreasonably long exposure age of 100 Ma at the find site using a production rate of 21 atoms of $^{21}Ne$ per g of $SiO_2$ per year (Niedermann, 2000), corrected for latitude and altitude of the find site (Stone, 2000) and a concentration of $^{21}Ne$-producing target elements (Na, Al, Si, Mg) at the upper end of the range given by Kramers et al. (2013). Lower concentrations of these elements result in even higher terrestrial exposure ages. The $^{21}Ne$ excess is thus interpreted to have been produced uniquely by cosmic-rays in an extra-terrestrial environment. It corresponds to nominal cosmic-ray exposure (CRE) ages of 0.004 to 0.09 Ma. This age is a rough estimate based on the production rate model given in (Leya and Masarik, 2009), assuming a meteoroid radius < 3 m, and a concentration of $^{21}Ne$-producing target elements (Na, Al, Si, Mg) at the upper end of the range given by Kramers et al. (2013), and no terrestrial contamination in any of these elements.

The original concentration of $^{21}Ne$-producing target elements in Hypatia is, however, difficult to assess. Kramers et al. (2013) suggested that most of the oxygen and magnesium in Hypatia might be the result of secondary (i.e. terrestrial) encrustations and fracture fillings. Such terrestrial contamination would imply that the above range of $^{21}Ne$ exposure ages would be a lower limit to the real exposure age of Hypatia. In contrast to $^{21}Ne$, cosmogenic $^{3}He$ is produced directly from C, the major element in Hypatia, essentially not affected by terrestrial contamination. A very conservative upper limit to the exposure age of Hypatia can therefore be calculated by assuming all measured $^{3}He$ is cosmogenic, in contrast to what we concluded in section 3.2.1. Again using the model calculations by Leya and Masarik (2009) with appropriate target element concentrations (dominated by C in this case), and no loss of cosmogenic $^{3}He$, the measured concentrations of $^{3}He$ (~ $4\text{-}11\times10^{-14}$ mol/g) correspond to CRE ages of 0.05 to 0.25 Ma in meteoroids with a radius < 3 m. This range is also low, similar to the nominal $^{21}Ne$ age range found above, in particular if we consider that likely in none of our samples more than perhaps 25% of the measured $^{3}He$ is actually cosmogenic (section 3.2.1). This suggests that Hypatia was either exposed to cosmic-rays in space for only a very short time, or was instead part of a significantly larger object. The slight excess of $^{3}He/^{4}He$ relative to He-Q could, thus, easily be explained by the addition of some cosmogenic $^{3}He$.

### 3.2.3 Argon

Except for some very low temperature steps, $^{40}Ar/^{36}Ar$ ratios in all Hypatia samples are significantly lower than the atmospheric ratio with values as low as 0.23 ± 0.38 measured in each of the three $CO_2$ laser extractions (H-N1, H-N2 & H-N3). Figure 2 shows the evolution of $^{40}Ar/^{36}Ar$ as a function of $^{36}Ar$ released in sample H-N1. Table 2 and Fig. S5 (in supplementary material) show that $^{40}Ar/^{36}Ar$ ratios as low as 1±4 and 0.6±0.3 in individual steps were also observed in samples H-N2 and H-N3 respectively. $^{40}Ar/^{36}Ar$ ratios of the total gas released during the $CO_2$ laser extractions are also very low with values of 3.5±1.7, 8.7±0.2 and 3.0±0.1 for samples H-N1, H-N2, and H-N3, respectively. For the samples analyzed in Zürich, $^{40}Ar/^{36}Ar$ ratios are less well constrained due to sizeable blank corrections, but in samples H-Z3 and H-Z4 the values are clearly much below the atmospheric ratio with an upper limit of 4 in sample H-Z4.

As already noted by Kramers *et al.* (2013), $^{40}Ar/^{36}Ar$ ratios below the atmospheric value provide clear evidence that Hypatia is extra-terrestrial. Values for the $^{38}Ar/^{36}Ar$ ratio are consistent both with the isotopic composition of Ar in phase Q (0.1873, (Busemann et al., 2000)) and with the atmospheric ratio. However, the very low $^{40}Ar/^{36}Ar$ values demonstrate that atmospheric Ar can only be a minor component.

### 3.2.4 Xenon and krypton

Xe and Kr were measured in sample H-N4 (8.4 mg) by stepwise extraction in an induction furnace. Data are shown in Table S1. Figure 3 shows the isotopic spectra of Xe released in the lowest temperature (400°C) and the highest temperature (1800-2200°C) steps, respectively. Xe released at low temperature is very similar to present-day atmospheric Xe and, therefore likely results from adsorbed atmospheric Xe. In contrast, the highest temperature step (Fig. 3b) released essentially pure Xe-Q (note that scales on Fig. 3a and 3b are distinct) plus a clear excess of radiogenic $^{129}$Xe. Xe data presented here are in stark contrast to those reported by Kramers *et al.* (2013) where no excess radiogenic $^{129}$Xe was found and large errors on isotopic ratios prevented any clear distinction between an extra-terrestrial component (e.g. Xe-Q) and the atmosphere. This is also illustrated in the three-isotope diagram in Fig. 4. Our data are plotted together with a selection of extreme data reported by (Kramers et al., 2013) characterized by particularly low $^{129}$Xe/$^{132}$Xe and $^{136}$Xe/$^{132}$Xe values. These data were used to propose the presence of the exotic Xe-G component, which falls near the origin in Fig. 4. Our data, however, do not show any hint for the presence of this Xe-G component in Hypatia.

The Kr data from analysis of H-N4 allow a less clear distinction between atmospheric Kr and Kr-Q. However, the Kr released in the high temperature steps is also fully consistent with the Kr-Q composition. Figure S6 (see supplementary material) is a three-isotope plot ($^{83}$Kr/$^{84}$Kr vs. $^{82}$Kr/$^{84}$Kr) showing some data from our study together with a selection of extreme data points from Kramers et al. (2013). Again, there is no evidence from our data for the presence of exotic Kr-G in Hypatia.

### 3.2.5 Nitrogen and carbon

Results for the two bulk samples analyzed in Paris are very reproducible, showing that these pre-combusted carbon-rich samples were composed of ca. 95% diamond, with identical $\delta^{15}$N-values of -100±1 ‰ (2σ) and low N-content (i.e. N/C-ratio) of 18 ppm. $\delta^{13}$C-values have an error-weighted mean of -3.44±0.14‰ (2 σ), a value in agreement with the result for the least contaminated samples of Kramers et al. (2013). Nitrogen data in all 3 stepwise $CO_2$ laser extractions conducted in Nancy are listed in Table 2. The evolution of the isotopic composition of nitrogen released during stepwise extraction of H-N3 is shown in Fig. 5 and is also representative for the two other extractions (see Figure S7 in supplementary material for the isotopic composition of $N_2$ in H-N1 and H-N2). In all samples, the major fraction (> 85 %) of nitrogen is released at mid- to high temperature. Remarkably, all three Hypatia samples released a very light component with an essentially constant $\delta^{15}$N around -110‰ over up to five high temperature steps. At lower temperatures, presumably some adsorbed atmospheric N ($\delta^{15}N$ = 0 ‰) was released, but it seems very likely that in at least some of these steps a second indigenous component with a positive $\delta^{15}N$ is also present. This component most clearly reveals itself in the first step of H-N3 (Fig. 5) with a $\delta^{15}N$ value of +25 ± 3 ‰ and in the second steps of H-N1 and H-N2. Since these steps may also have been affected by some adsorbed atmospheric nitrogen, we conclude that the indigenous low-T component in Hypatia has a $\delta^{15}N$ ratio around +25‰ or higher. As discussed below, such high $\delta^{15}N$ values are common in meteorites and rare on Earth. Remarkably, total $\delta^{15}N$ values of nitrogen extracted from samples in Nancy (e.g. -95 ± 4 ‰ (2σ) in sample H-N1; Table 2) agree well with values obtained in Paris on Hypatia diamonds ($\delta^{15}N$ = -100 ± 1 ‰ (2σ)). The slightly higher $\delta^{15}N$ value of -86±1 ‰ (1σ) for sample H-N3 (Table 2) is likely due to the presence of the heavy component released during the first heating step in Nancy and probably removed by the pre-combustion at 600°C in Paris. For the subsequent detailed discussion, Fig. 5 shows ranges of $\delta^{15}N$ values observed in different meteoritic materials. We note that nitrogen components similar to both the heavy and the light nitrogen found in Hypatia have been detected in various extra-terrestrial samples.

### 3.2.6 Concentrations and elemental ratios

The relative elemental abundances of noble gases and nitrogen can be diagnostic of their carrier phases (Ott, 2014). Figure 6 shows the mean (thick black line) and the ranges (grey area) of total concentrations of noble gases and nitrogen extracted from Hypatia samples analyzed in this study. Concentrations in the different samples are quite uniform, e.g. vary by less than a factor of 2.5 for $^{36}Ar$ ($3.2x10^{-11}$ mol/g). Volatile element abundances in bulk ureilites, a bulk Almahata Sitta ureilitic fragment (Murty et al., 2010) and bulk Goalpara (Göbel et al., 1978) ureilite, in carbon-rich residues extracted from ureilites (Rai et al., 2003a), in a graphite inclusion from the Canyon Diablo iron meteorite (Matsuda et al., 2005), in bulk Monument Draw acapulcoite (McCoy et al., 1996) and in the Earth's atmosphere (in mol per g of atmosphere) are also shown for comparison. In our study and in references listed above all abundances are the results of total extractions and may include in some cases an atmospheric component. Concentrations of noble gases and nitrogen in Hypatia samples are considerably lower than the high values measured in phase Q. For example, $^{132}Xe$ concentrations in phase Q reach values up to $1.5x10^{-11}$ mol/g (Busemann et al., 2000) versus $4.3x10^{-14}$ mol/g in Hypatia. By contrast, the abundance levels and elemental pattern in Hypatia is similar to that of bulk ureilites and also of the Canyon Diablo graphite nodule, apart from He that seems enriched relative to the graphite nodule with a mean value of $4.7\times10^{-10}$ mol/g. The abundance pattern in the Monument Draw acapulcoite is also broadly compatible with that of Hypatia for He, Ne, and Xe. Hypatia shows much lower volatile concentrations than carbon-rich residues extracted from ureilites despite the fact that Hypatia is mainly composed of carbon. These low concentrations prevent us from directly linking Hypatia to the carbon-rich part of ureilites (see discussion section 4.2).

**4. Discussion**

The data presented here undoubtedly confirm and strengthen the conclusion by Kramers et al. (2013) that Hypatia is an extraterrestrial material. Apart from the confirmation of very low $^{40}Ar/^{36}Ar$ ratios, this is also clearly shown by the isotopic composition of He, Ne, and Xe, which in many extraction steps and bulk samples is essentially identical to the Q component, ubiquitous in many meteorite classes. Furthermore, given that values of $\delta^{15}N$ below -40 ‰ have never been reported in terrestrial samples and that the nitrogen isotopic composition in Hypatia is similar to values found in different type of meteoritic materials, a $\delta^{15}N$ value of -110‰ for more than 85% of the N budget (Table 2) is a new convincing piece of evidence for the extraterrestrial nature of Hypatia. Elemental abundances of the ultravolatile elements analyzed here are also very different from the terrestrial atmospheric pattern and are in the range observed for different meteoritic samples (Fig. 6). Furthermore terrestrial impact diamonds occur in a geological context where graphite-bearing target rocks are present such as gneisses in the Ries, Popigai, and Lappajärvi craters (Langenhorst et al., 1999). In case of Hypatia, the potential target is made of Nubian sandstone. The absence of graphite-bearing target rock in this area further supports the conclusions that Hypatia must be extraterrestrial.

Although the indications for ‘exotic’ extrasolar components reported by Kramers et al. (2013), discussed below, are tenuous, there are further, quite robust differences between the noble gas data obtained by these authors and the present dataset, and also one marked similarity. (i) Their data for Ar, Kr, and Xe are dominated by terrestrial atmospheric contamination, which persisted to high temperatures in the degassing experiments. (ii) They found no excess $^{129}Xe$. (iii) While their inferred abundances of the extraterrestrial noble gas components for He, Ar, and Xe are 5 to 10 times lower than those determined in the present study, their corresponding He/Xe and Ar/Xe ratios are very similar to our data (and significantly higher than those in typical Q gas). The grains analyzed by Kramers et al. (2013) came from a different subsample of Hypatia to ours, and it is thus likely that the stone is heterogeneous on a mm to cm scale with regard to its iodine and trapped noble gas abundances, but homogeneous with regard to the composition of its dominant extraterrestrial noble gas component.

In the following discussion we will explore possible links between Hypatia and extra-terrestrial objects from which it may originate based on the findings presented above.

**4.1 A cometary origin for Hypatia?**

Our study does not confirm the presence of any G noble gas component as reported by Kramers et al. (2013). The presence of "exotic" G-gases, normally found in presolar SiC grains, could have been a clue to a cometary origin of Hypatia under the premise that cometary matter should be rich in primordial noble gas components produced by nucleosynthesis in stars. In further contrast to Kramers et al. (2013) we unequivocally show that the analyzed fragments of Hypatia are rich in isotopically “normal” Q gases, a major primordial noble gas component in many different meteorite classes originating from the asteroid belt. Furthermore, only little is known about volatile elements in comets and for example Ne isotopically similar to Ne-Q in chondrites has been measured in Stardust samples (Marty et al., 2008).

**4.2 Possible links of Hypatia with known carbon-rich meteoritic materials**

Because phase-Q noble gases are ubiquitous in many meteorite classes (Ott, 2014), nitrogen isotopic composition may be more useful for exploring possible links between Hypatia and known extraterrestrial objects, as N isotopic compositions are very variable in different extraterrestrial samples (Füri and Marty, 2015). On the right hand side of Fig. 5, $\delta^{15}N$ ranges of various C-rich phases in meteorites are shown. The very light ($\delta^{15}N \approx -110‰$) and reproducible nitrogen released from all three analyzed Hypatia samples at high temperature is reminiscent of a main component measured in ureilitic diamonds ($\delta^{15}N < -100‰$; (Rai et al., 2003b)). In addition, the likely minor N component released at low to mid temperatures, with $\delta^{15}N > 25$ ‰, is reminiscent of an isotopic signature of a component with $\delta^{15}N$ higher than 19 ‰ and reaching values up to 100 ‰ (Rai et al., 2003b; Yamamoto et al., 1998) released at low temperature from graphite in ureilites. It is worth noting that such values are released at low temperatures which is again consistent with our observations (Rai et al., 2003b).

Additionally, trapped noble gases with an isotopic composition similar to the Q component and with elemental abundances ratios more similar to those in Hypatia than in phase Q are present in ureilites (Rai et al., 2003a). Note here that the bulk $\delta^{13}C$ value of -3.4 ± 0.1 ‰ measured in diamond is also in the range of values measured in ureilites (Grady and Wright, 2003). Unlike in Hypatia, however, despite an active search, no excess $^{129}Xe$ from the decay of $^{129}I$ has ever been reported in ureilites (Göbel et al., 1978; Rai et al., 2003a) (Fig. 4). This latter observation suggests that Hypatia may not be connected to ureilites after all.

Grady and Wright (2003) found isotopically light N in the range of $\delta^{15}N$ = -50 to -20 ‰ in graphite nodules of iron meteorites. Such nodules are mainly found in IAB and IIICD types of meteorites (Benedix et al., 2000). While this range of $\delta^{15}N$ is not overlapping with that of the main N component in Hypatia (Fig. 5), C-rich samples of iron meteorites appear to contain nitrogen with quite heterogeneous $\delta^{15}N$ values. For example, isotopically very light nitrogen with $\delta^{15}N$ < -82 ‰ has been reported in Copiapo (IAB) (Ponganis and Marti, 2007). Notably, radiogenic $^{129}Xe$ as well as trapped noble gases similar to Q have also been found in graphite nodules of Canyon Diablo (IAB) (Matsuda et al., 2005). Furthermore Kramers et al. (2013) reported the presence in Hypatia of bright inclusions composed of a Fe-Ni-Cr alloy and troilite reminiscent of the major constituents of iron meteorites. The IAB iron meteorite ALH 77283 contains also troilite-graphite-schreibersite-cohenite inclusions rich in diamond-lonsdaleite nodules that may be similar to Hypatia (Clarke et al., 1981). Finally the bulk $\delta^{13}C$ value of -3.4±0.1 ‰ measured in this study is also in the range of -30 to +4 ‰ measured in graphite nodules in iron meteorites even if this value, given the large range measured among different samples, is not diagnostic of this type of extra-terrestrial material (Grady and Wright, 2003). Graphite rich parts of iron meteorites may thus appear linked to Hypatia, although the heterogeneity of the N isotopic composition in such samples prevents us to unequivocally interpret Hypatia as a graphite nodule from an iron meteorite.

Acapulco and Lodran, the type specimens of the primitive achondrite groups of acapulcoites and lodranites contain small µm-sized graphite nodules or carbonized veins with isotopically often very light nitrogen ($\delta^{15}N$ as low as -166 ‰ in Acapulco (Charon et al., 2014)). Furthermore, acapulcoites and lodranites also carry noble gases of Q-composition as well as excess $^{129}Xe$ (McCoy et al., 1997; Palme et al., 1981). Acapulcoites and lodranites may, therefore, also share a genetic link with Hypatia.

Hypatia is cm-sized (Barakat, 2012). Only graphite nodules in iron meteorites typically reach this size. Carbon-rich phases in other known meteoritic classes discussed here (ureilites, acapulcoites and lodranites) are much smaller (µm- to mm-sized) except in one case where a large cm-sized carbon-rich nodule, maybe containing diamonds, has been found in the Portales Valley (H6) chondrite (Ruzicka et al., 2000). If size matters to establish a link between Hypatia and other know extraterrestrial objects none of the candidates discussed here, except graphite nodules from iron meteorites, match this feature (Table 4).

In summary, while Hypatia clearly is a different type of material than any of the carbon-rich phases discussed in this section, its noble gas and nitrogen signatures share many characteristics with some of these phases, although none matches perfectly (see Table 4 for a comparison between the features met in Hypatia and those in known extraterrestrial carbon-rich phases). Therefore, it seems possible that the volatile inventory of Hypatia or its parent material is related to the volatiles in these carbon-rich phases of known meteorite classes and that all these carbon-rich lithologies may have sampled the same geochemical reservoir.

**4.3 A link between Hypatia and the Libyan Desert Glass?**

Kramers et al. (2013) suggested that Hypatia is a remnant of the impactor that created the Libyan Desert Glass. Irrespective of whether Hypatia is of cometary or asteroidal origin, their main argument for a causal relation between Hypatia and LDG was that a large object was required to generate diamonds by impact. The shock pressures required to produce impact diamonds from graphite must generally exceed about 25 - 30 GPa. In case of a terrestrial impact the projectile should be at least some meters in diameter. This would be consistent with the likely minimum size of Hypatia inferred from the cosmogenic noble gases as discussed further down in this paragraph. A crater or a crater strewn field would also result (Collins et al., 2005). Such pressures are indeed recorded in bedrocks of LDG, however a crater or a crater strewn field related to Hypatia is not known. It should also be noted that so far some 40 meteorites of different classes (named “Great Sand Sea n°XX ”), including ordinary chondrites, iron meteorites and lodranites, have been found near the LDG area. The find location is thus not a compelling argument for a causal relationship. The XRD and TEM studies presented here also do not allow us to directly decide whether the diamonds in Hypatia formed upon impact on Earth or by an earlier collision in space, for example during the event which ejected the Hypatia-bearing meteoroid from its parent body. However, nitrogen in Hypatia is not isotopically equilibrated between the low-to-medium temperature release of N with a $\delta^{15}N$ component around +25 ‰ or higher, and a medium-to-high temperature component depleted in $^{15}N$ ($\delta^{15}N$ values close to -110 ‰). Part of the nitrogen released in the first heating steps is probably of terrestrial origin, but likely not all since $\delta^{15}N$ values above 20 ‰ are extremely rare on Earth but common in extraterrestrial carbon-rich material, e.g. insoluble organic matter in carbonaceous chondrites or low to medium heating steps of ureilites (Rai et al., 2003b). The high temperature extraction of isotopically light N is consistent with its occurrence in diamond as N is the most abundant impurity found in diamonds and its substitution in the place of carbon explains the difficulty, and thus the high-temperature required, to release it from the diamond structure (Kaiser and Bond, 1959). Hence, the isotopic disequilibrium of N among different carbon-rich phases implies that, in Hypatia, diamonds are unlikely to have formed by shock of the C-rich material during atmospheric entry. In the parent meteoroid, diamonds hosting light N likely co-existed with non-diamond, C-rich phases hosting heavy N before encounter with Earth, as observed for instance in ureilitic C-rich veins. Whatever the processes involved, the dichotomy in $\delta^{15}N$ between diamond-rich and amorphous carbon-rich lithologies must have developed in space otherwise the impact and the production of diamond would have homogenised the isotopic composition of nitrogen. Thus, the disequilibrium in nitrogen isotopic compositions argues against production upon impact on Earth. The very low nominal noble gas exposure ages of Hypatia may also be considered in this context. Because the $^{22}Ne/^{21}Ne$ ratio of the cosmogenic component could not be determined due to the dominant presence of trapped Ne and because so far no cosmogenic radionuclide data for Hypatia is available, information about the shielding of Hypatia during its journey towards Earth (i.e. the size of the Hypatia parent meteoroid and the preatmospheric depth of Hypatia within that meteoroid) is unconstrained. The exposure age of Hypatia on the order of a mere 100,000 years was therefore estimated by assuming a production rate valid for a "typical" meteoroid size < 3 m. Hence, the qualification of this number as "nominal exposure age". True meteorite exposure ages considerably less than a million years are very rare. If Hypatia had been brought to Earth as part of an iron meteorite, its exposure age would be expected to be at least 10-20 Ma, but more likely to be longer than 100 Ma (Herzog and Caffee, 2014). Also among stony meteorites, exposure ages of less than 1 Ma are only found for some types of carbonaceous chondrites (Nishiizumi and Welten, 2005), with most other ages being at least a few Ma and ranging up to some 100 Ma. Thus, it seems likely that the low nominal exposure age of Hypatia actually is the result of a much larger shielding, and, hence, larger meteoroid size, than assumed above.

While no clear-cut conclusion on the size of the Hypatia parent meteorite can be drawn from these considerations, it seems rather likely that it was a body of at least a few meters in diameter. On the other hand, if the parent meteorite/asteroid had been even considerably larger, as would be implied by a supposed connection with the LDG event, then Hypatia would have to originate from the outermost few meters of this body. Otherwise, the low but measurable concentrations of cosmogenic $^{21}$Ne could not have been produced even during billions of years of cosmic ray exposure. The example of the Canyon Diablo iron meteorite shows that surviving large chunks of a large bolide are indeed likely to originate from the rear near-surface portions of the impacting projectile (Bjork, 1961). In summary, the concentration of cosmogenic $^{21}$Ne in Hypatia hints at a parent object of at least a few meters in diameter, but not necessarily large enough to have been able to create the LDG or to produce the shock-diamonds upon impact on Earth. Nevertheless, the suggestion by Kramers et al. (2013), that more fragments similar to the Hypatia stone might be found near the site, is further supported by the low cosmogenic $^{21}$Ne concentration.

## 5. Summary and Conclusions

The analyses presented here confirm conclusions by Kramers et al. (2013) that the enigmatic pebble Hypatia represents an unusual type of extraterrestrial material. In addition to the clear cut evidence pointed out by Kramers *et al.* (2013) (e.g. $^{40}$Ar/$^{36}$Ar ratios below the atmospheric value), our study shows: $^{3}$He/$^{4}$He, $^{20}$Ne/$^{22}$Ne, Xe isotopes, and likely Kr and Ar close to the isotopic composition of the ubiquitous component phase Q in meteorites, isotopic composition of nitrogen similar to components found in ureilites and maybe in graphite inclusions of iron meteorites or acapulcoites, and finally small concentrations of cosmogenic $^{21}$Ne produced in space. We also found significant differences from data previously published on Hypatia. In particular we found no evidence for the presence of any presolar signature (e.g. the G component) that was used to infer a cometary origin for Hypatia. At this time, we cannot definitively associate Hypatia with any known type of meteorites as each candidate considered here (ureilites, graphite nodules in iron meteorites or in acapulcoites/lodranites) fails to reproduce all features of the ultravolatile elements determined in this study (Table 4). However, from our comparisons it appears that Hypatia may be related to differentiated cosmochemical objects and thus might present a great opportunity to understand the origin and mode of survival of primordial noble gases and nitrogen in such objects (Wieler et al., 2006).

**Acknowledgments**

Yves Marrocchi, David Bekaert, Marc Chaussidon and Léo Martin are gratefully acknowledged for fruitful discussions. We appreciate the constructive comments by two anonymous reviewers. This study was funded in Nancy (France) by the European Research Council under the European Community's Seventh Framework Programme (FP7/2007-2013 grant agreement no. 267255 to B.M.). M. M. is supported by an Ambizione grant (project PZ00P2_154874) from the Swiss National Science Foundation. Further financial support was provided by the Deutsche Forschungsgemeinschaft (LA 830/14-1 to FL). This is CRPG contribution #2377.

**Figure captions**

Figure 1: Isotopic composition of neon in Hypatia samples. Empty symbols are for individual heating steps and filled symbols for bulk samples. The isotopic compositions of Ne in Earth's atmosphere, Ne-HL, Ne-Q and Ne-G are shown as references (Ott, 2014). Arrow shows the tendency toward the cosmogenic component ($^{21}Ne/^{22}Ne \approx 0.9$; (Leya et al., 2001)). Errors at 1σ.

Figure 2: Evolution of the $^{40}Ar/^{36}Ar$ ratio during step-heating of the H-N1 sample. Note the log-scale for the y axis showing very low ratios well below the atmospheric value of 298.56 (Lee et al., 2006). Ranges correspond to 1σ.

Figure 3: Isotopic spectra of Xe released at 400°C (a) and between 1800 and 2200°C (b) during the induction furnace experiment. The isotopic composition is expressed with the delta notation relative to the isotopic composition of Xe in Solar Wind (Meshik et al., 2014) ($\delta^{i}Xe = \left(\left({}^{i}Xe/{}^{132}Xe\right)_{mes} / \left({}^{i}Xe/{}^{132}Xe\right)_{SW} - 1\right) \times 1000$). Grey triangles in (a) represent the isotopic composition of Earth's atmosphere and grey triangles in (b) represent the isotopic composition of Xe in phase Q (Busemann et al., 2000). Errors at 2σ.

Figure 4: Three-isotope plot of xenon. Small black-filled circles are for individual heating steps, the large black-filled circle corresponds to the total of all extraction steps. Reference values of Solar Wind (SW) xenon, Xe-Q, G-Xe, and atmospheric Xe are shown for comparison (Ott, 2014) as well as the mean value of the $^{129}Xe/^{132}Xe$ ratio measured in ureilites (Rai et al., 2003a). A representative set of data by (Kramers et al., 2013) are shown with small grey axes and error bars. Error bars are at 2σ.

Figure 5: Ranges of isotopic compositions of nitrogen released during step-heating of the H-N3 sample. Ranges for individual steps correspond to errors at 1σ.

Figure 6: Range of elemental abundances (in mol/g) of volatile elements in Hypatia compared to several types of meteorites. The range for $^{4}He$ abundances is from the analysis of H-Z1 to H-Z5 assuming that all $^{4}He$ is trapped. The range for $^{20}Ne$ (trapped) abundances is from H-Z2 to H-Z5 and from H-N1 to H-N3. Value from H-Z1 analysis is rejected because it is unrealistically low due to high corrections due to the presence of water. Range for $^{36}Ar$ is from H-Z1 to H-Z5 and from H-N1 to H-N3. Range for $N_2$ is from H-N1 to H-N3. Range for $^{84}Kr$ is from the unique analysis on sample H-N4 and the range corresponds to a relative high uncertainty due to a perfectible calibration of the Kr content of the standard bottle. However,

it does not change any conclusion of this study. Only the value of $^{132}Xe$ amount obtained during analysis of sample H-N4 is shown. See text for details about references.

**Table captions**

Table 1: Analytical techniques and list of chemical species analyzed in Hypatia samples during this study.

Table 2: Data of Ne, Ar, and $N_2$ measured in Hypatia samples H-N1 to H-N3. Errors at 1$\sigma$.

Table 3: Data of He, Ne, and Ar measured in Hypatia samples H-Z1 to H-Z5. Errors at 1$\sigma$.

Table 4: C-rich meteoritic lithologies that partially match the features observed in Hypatia. The "✓" symbol denotes a match with a feature, "✖" a dismatch and "-" the difficulty to conclude. See text for details and references.

## Figures

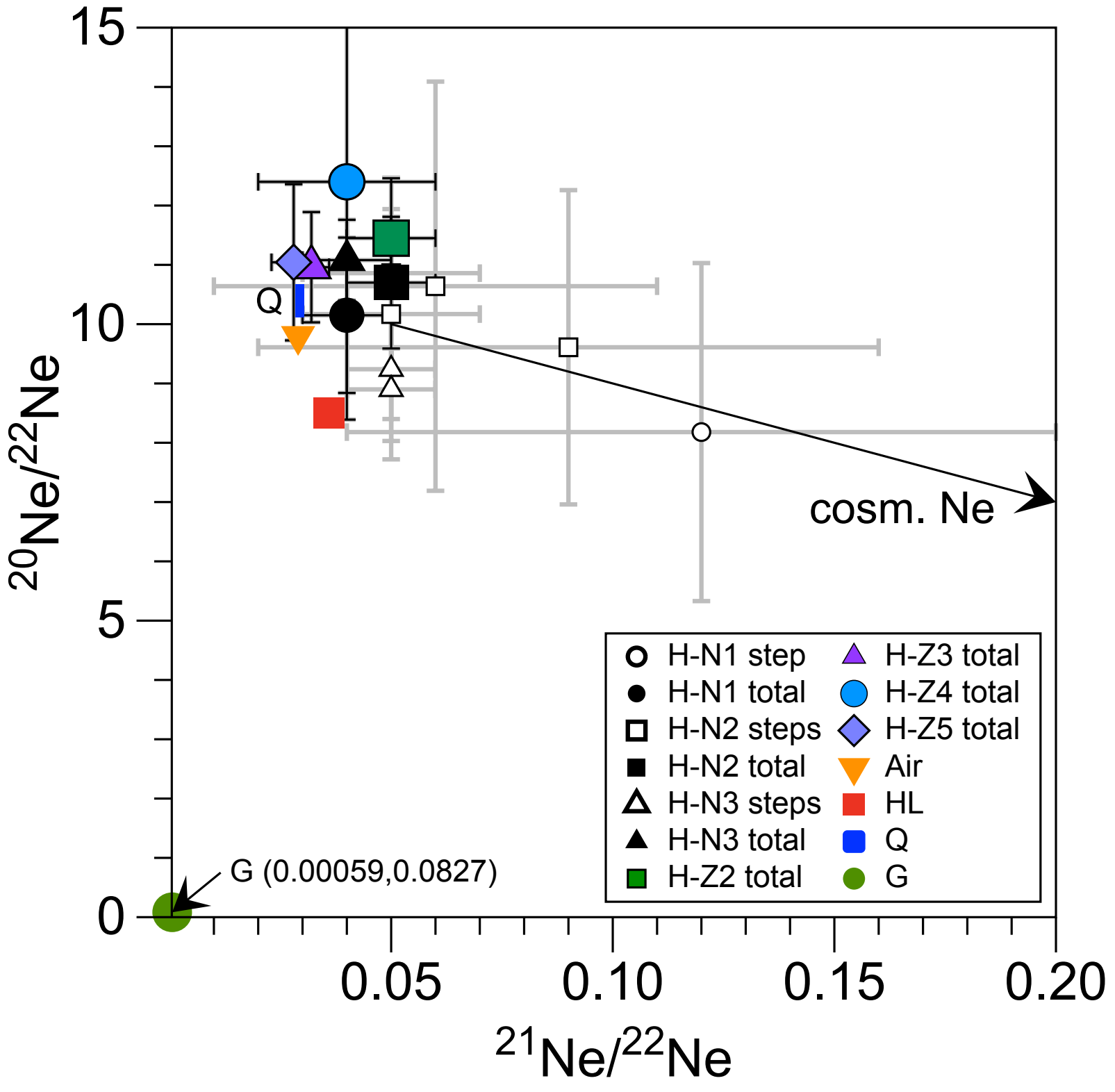

Figure 1

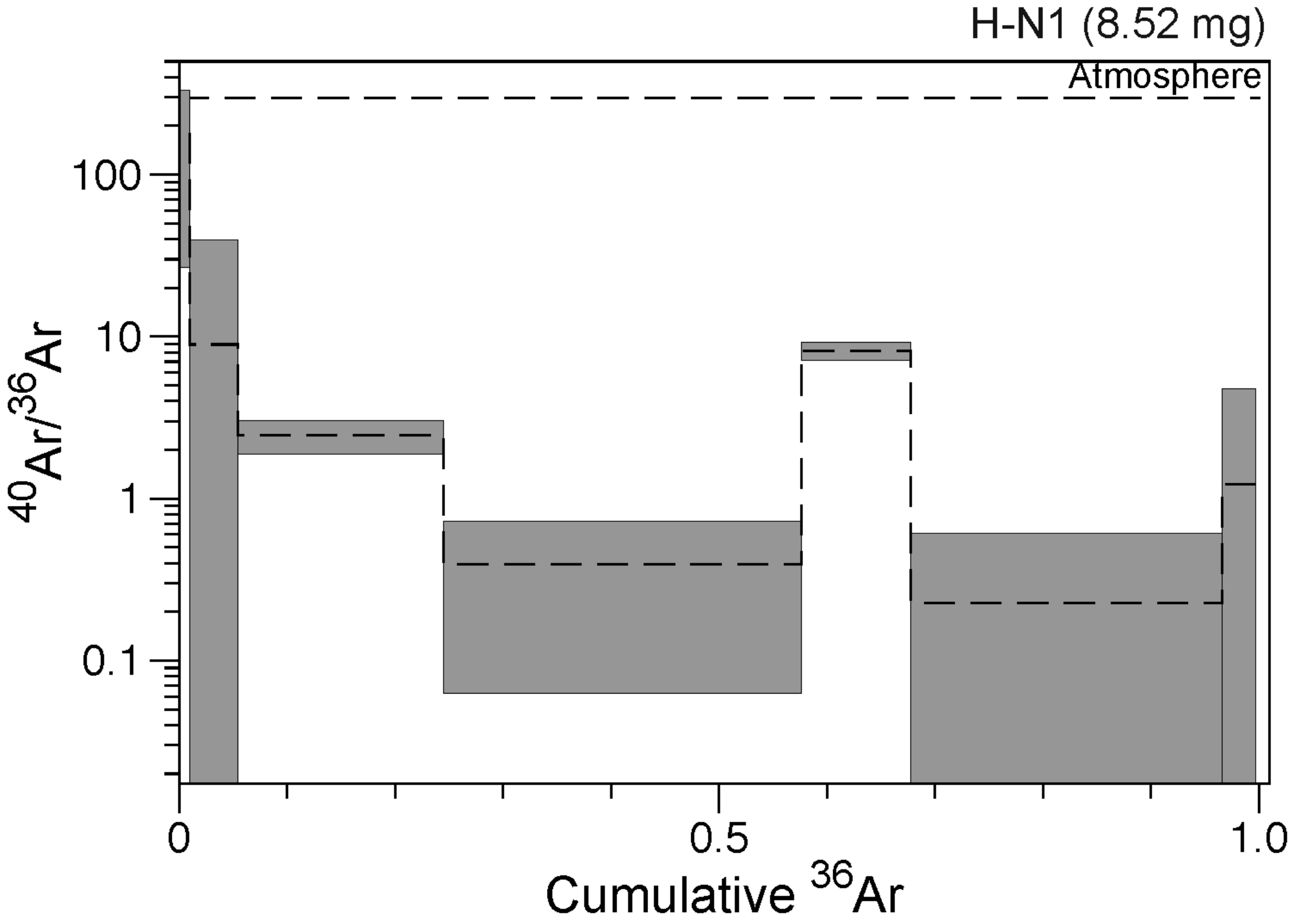

Figure 2

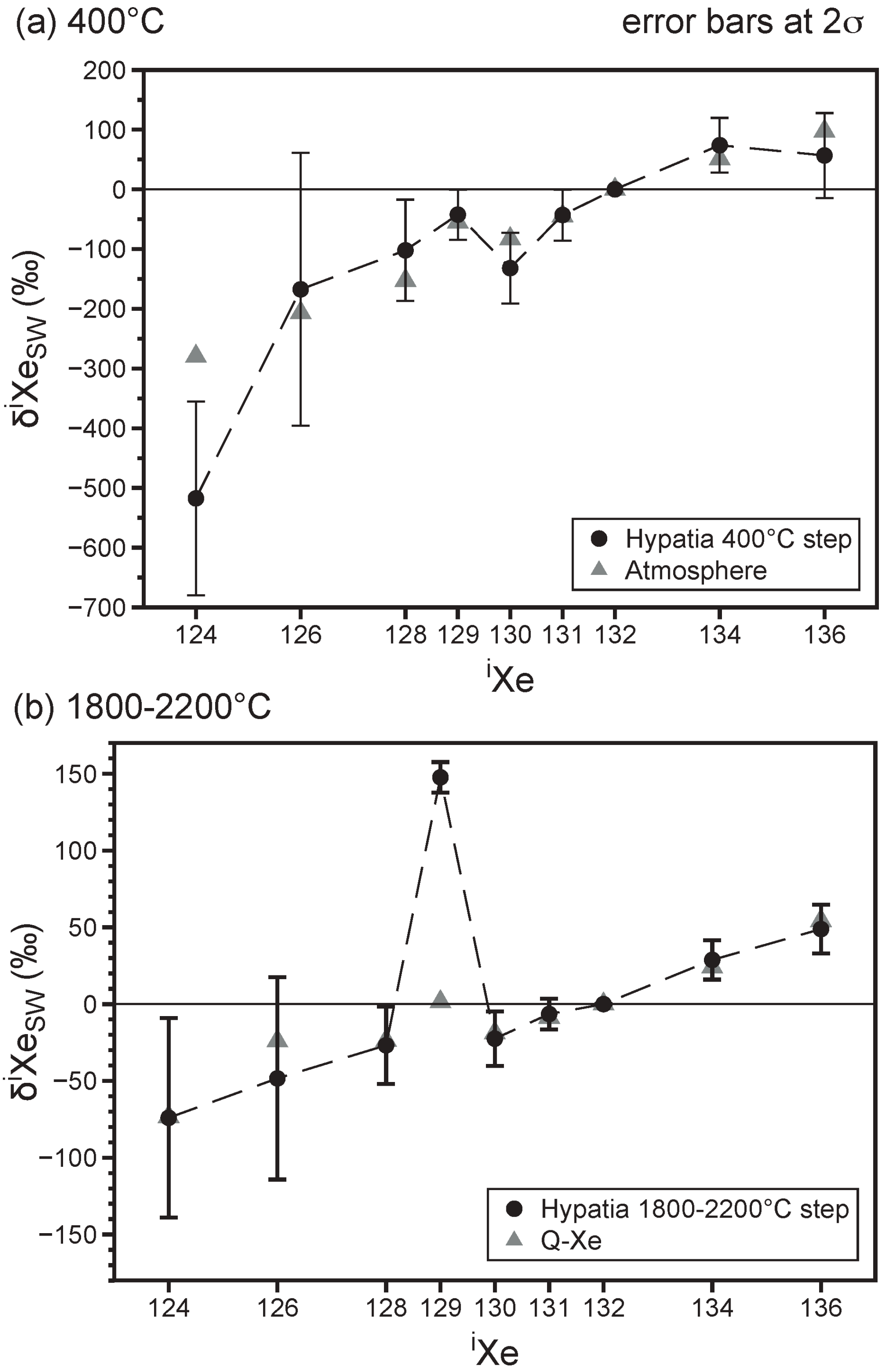

Figure 3

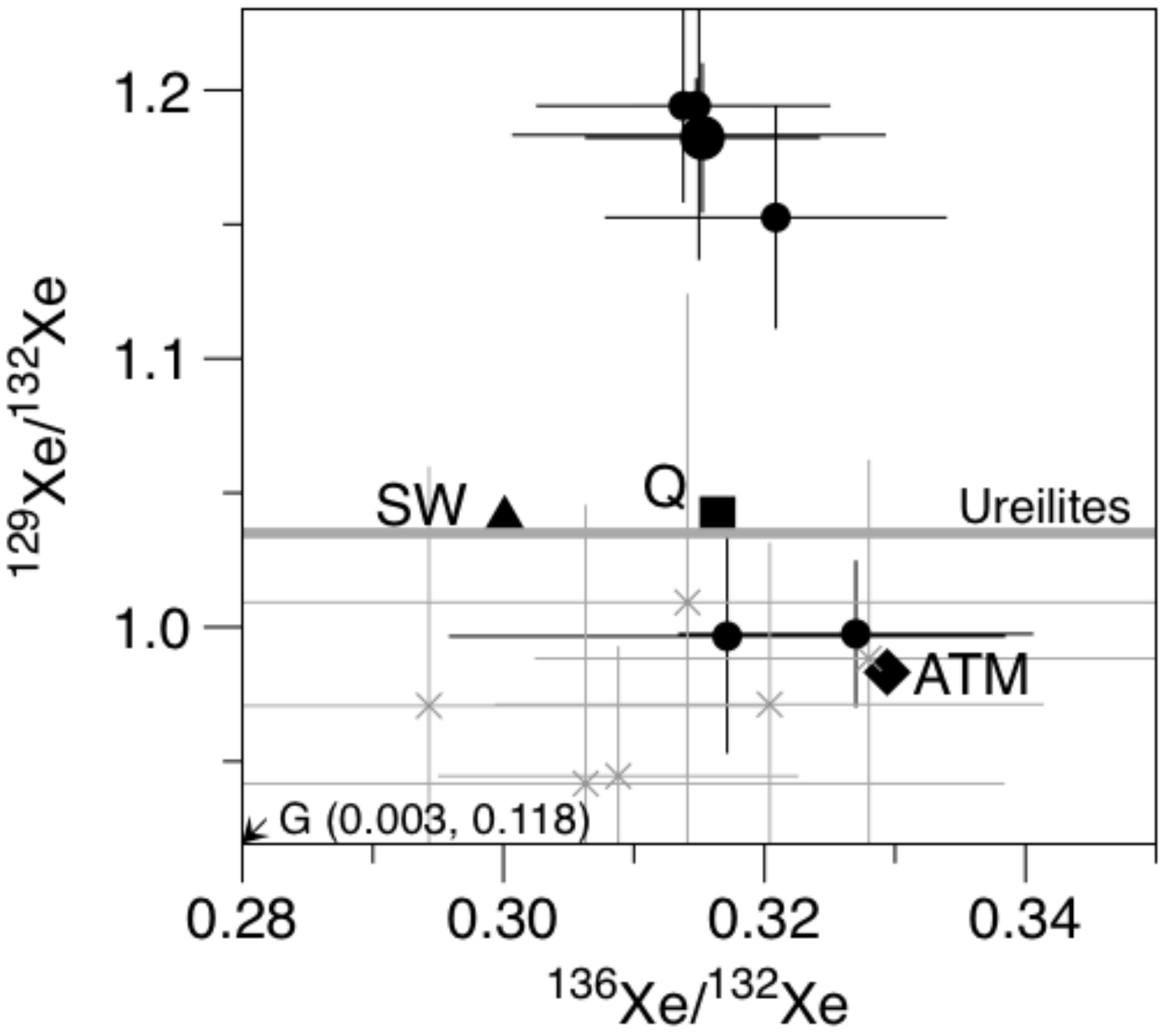

Figure 4

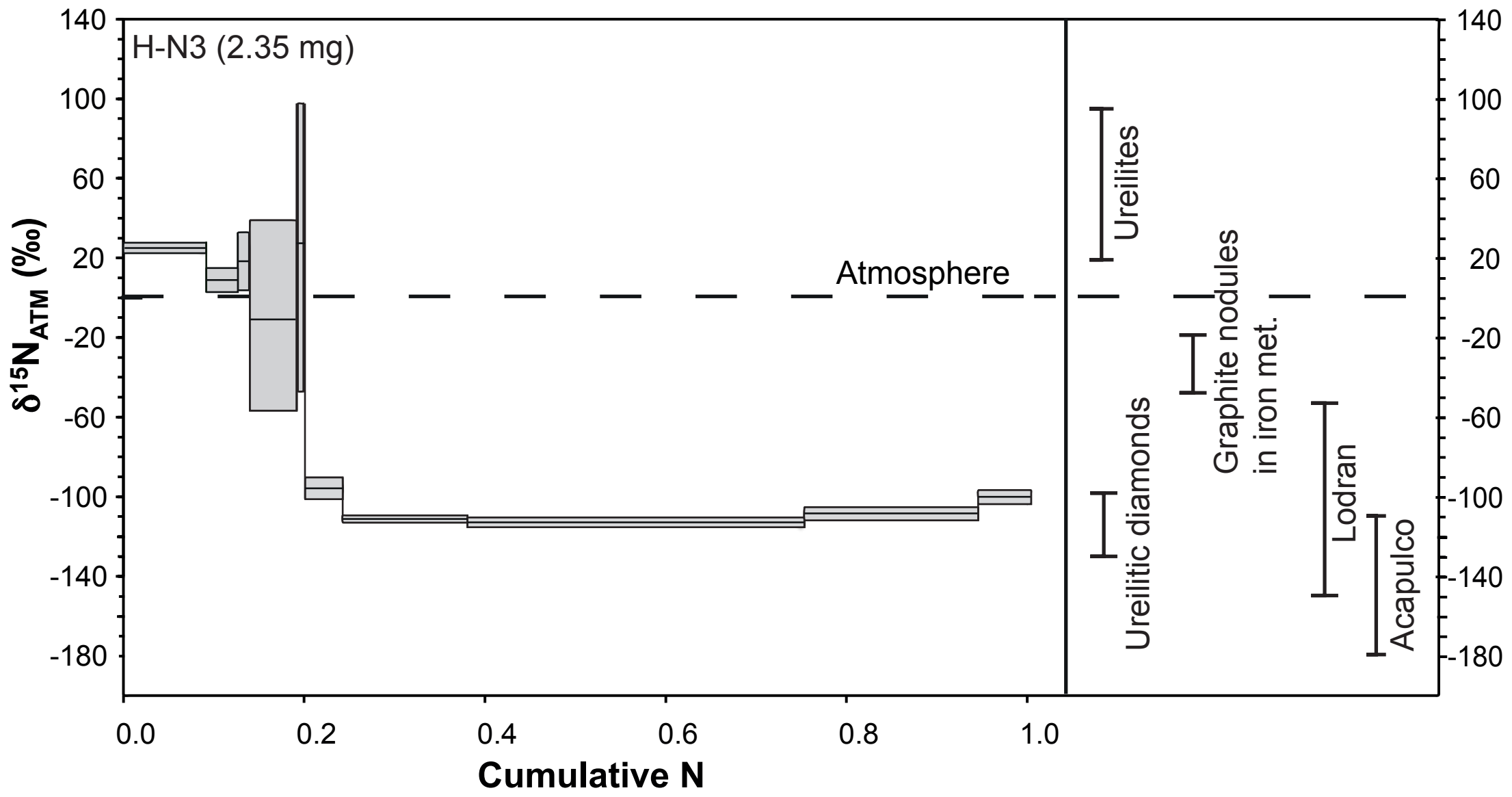

Figure 5

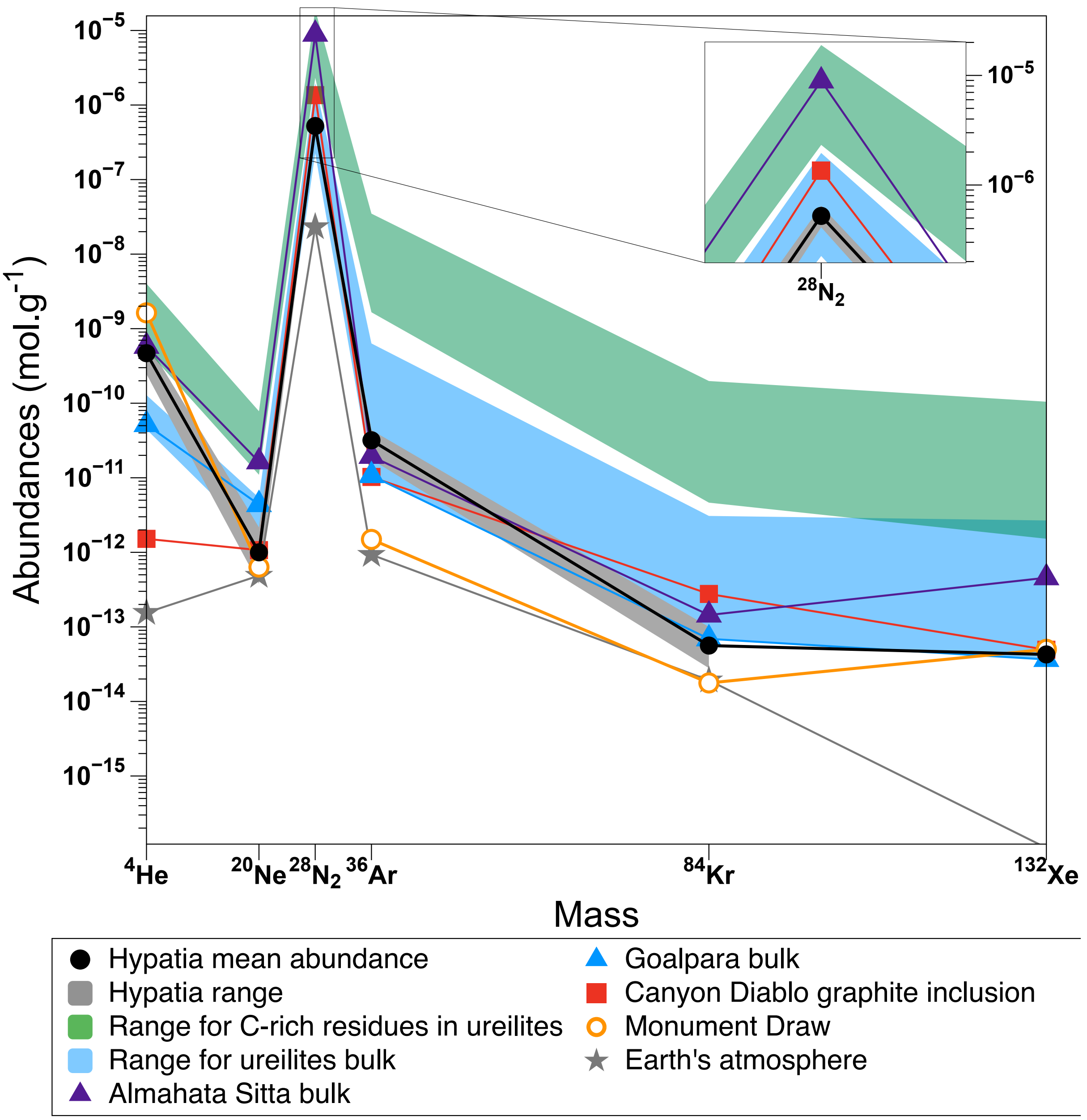

Figure 6

**Table 1**

| Samples Id. | Mass (mg) | Analyzed species | Extraction method |
|---|---|---|---|
| H-N1 | 8.52 | $N_2$, Ne, Ar | stepwise CO2 laser extraction |
| H-N2 | 1.07 | " | " |
| H-N3 | 2.35 | " | " |
| H-N4 | 8.4 | Ar, Kr, Xe | stepwise induction furnace extraction (400-2200°C) |
| H-Z1 | 1.0 | He, Ne, Ar | single step IR laser extraction |
| H-Z2 | 2.9 | " | " |
| H-Z3 | 2.9 | " | single step furnace extraction |
| H-Z4 | 1.0 | " | " |
| H-Z5 | 2.8 | " | " |
| H-P1 | 1.6 | $N_2$, C | on-line combustion |
| H-P2 | 1.6 | $N_2$, C | " |

Samples analysed in Nancy labelled "H-N…", samples analysed in Zürich labelled "H-Z…", samples analysed in Paris labelled "H-P…" In all cases, all stable isotopes of the listed species were analysed, although in some cases large blank corrections inhibited a meaningful determination of gas amounts and/or isotopic ratios (see Results section).

# Table 2

| Sample | Extraction step | $^{22}Ne$ (mol/g) | ± | $^{20}Ne/^{22}Ne$ | ± | $^{21}Ne/^{22}Ne$ | ± | $^{36}Ar$ (mol/g) | ± | $^{40}Ar/^{36}Ar$ | ± | $^{38}Ar/^{36}Ar$ | ± | $N_2$ (mol/g) | ± | $\delta^{15}N$ (‰) | ± |
|---|---|---|---|---|---|---|---|---|---|---|---|---|---|---|---|---|---|
| H-N1 (8.52 mg) | #1 | 1.4E-14 | 3.E-15 | 9.7 | 0.8 | 0.071 | 0.011 | < blk | | n.d. | | n.d. | | 2.23E-08 | 2.E-10 | -9 | 57 |
| | #2 | < blk | | n.d. | | n.d. | | 4.324E-13 | 9.4E-14 | 178.3 | 151.5 | 0.244 | 0.085 | 2.97E-08 | 2.E-10 | 14 | 8 |
| | #3 | 1.3E-14 | 3.E-15 | n.d. | | 0.084 | 0.015 | < blk | | n.d. | | n.d. | | 5.19E-09 | 2.E-10 | 2 | 40 |
| | #4 | 1.1E-14 | 2.E-15 | 9.9 | 0.9 | 0.015 | 0.009 | 1.954E-12 | 5.7E-14 | 8.9 | 30.8 | 0.205 | 0.016 | 1.48E-08 | 2.E-10 | 9 | 14 |
| | #5 | 4.9E-14 | 1.E-14 | 9.1 | 0.6 | 0.033 | 0.011 | 8.321E-12 | 5.1E-14 | 2.4 | 0.6 | 0.187 | 0.001 | 8.00E-08 | 3.E-10 | -111 | 3 |
| | #6 | 4.0E-14 | 5.E-15 | 8.2 | 0.6 | 0.011 | 0.008 | 1.450E-11 | 8.7E-14 | 0.4 | 0.3 | 0.185 | 0.001 | 2.16E-07 | 8.E-10 | -112 | 1 |
| | #7 | 2.1E-14 | 3.E-15 | 10.5 | 0.6 | 0.037 | 0.019 | 4.416E-12 | 3.3E-14 | 8.2 | 1.1 | 0.185 | 0.002 | 5.75E-08 | 3.E-10 | -107 | 5 |
| | #8 | 3.9E-14 | 2.E-15 | 8.9 | 0.3 | 0.010 | 0.008 | 1.260E-11 | 8.6E-14 | 0.2 | 0.4 | 0.185 | 0.002 | 2.08E-07 | 7.E-10 | -105 | 2 |
| | #9 | 1.1E-14 | 2.E-15 | 9.7 | 0.9 | 0.056 | 0.013 | 1.352E-12 | 2.0E-14 | 1.2 | 3.5 | 0.194 | 0.005 | 1.09E-08 | 2.E-10 | -65 | 23 |
| | #10 | 1.6E-14 | 2.E-15 | 8.2 | 0.6 | 0.116 | 0.023 | 1.408E-13 | 7.7E-15 | n.d. | | 0.173 | 0.024 | 1.48E-09 | 2.E-10 | 116 | 145 |
| | Total | 2.1E-13 | 3.E-14 | 10.1 | 1.3 | 0.041 | 0.011 | 4.371E-11 | 1.8E-13 | 3.5 | 1.7 | 0.187 | 0.001 | 6.46E-07 | 1.E-09 | -95 | 2 |
| H-N2 (1.07 mg) | #1 | 6.2E-15 | 2.E-15 | n.d. | | n.d. | | 6.897E-13 | 1.0E-14 | 198.7 | 6.3 | 0.180 | 0.006 | 7.23E-08 | 3.E-10 | -29 | 46 |
| | #2 | 1.11E-13 | 3.E-15 | n.d. | | n.d. | | 1.474E-13 | 5.1E-15 | 5.8 | 25.9 | 0.169 | 0.019 | 6.37E-09 | 2.E-10 | 40 | 25 |
| | #3 | 2.06E-13 | 4.E-15 | n.d. | | n.d. | | 1.453E-12 | 1.6E-14 | n.d. | | 0.188 | 0.004 | 1.70E-08 | 2.E-10 | 1 | 11 |
| | #4 | 3.6E-14 | 2.E-15 | 9.8 | 0.4 | 0.037 | 0.005 | 6.394E-12 | 5.2E-14 | 22.8 | 0.6 | 0.183 | 0.001 | 4.04E-08 | 2.E-10 | -107 | 4 |
| | #5 | 2.8E-14 | 2.E-15 | 10.9 | 0.5 | 0.048 | 0.005 | 2.125E-11 | 1.4E-13 | n.d. | | 0.188 | 0.001 | 2.17E-07 | 8.E-10 | -111 | 2 |
| | #6 | 2.3E-14 | 2.E-15 | 7.3 | 0.4 | 0.031 | 0.004 | 3.035E-12 | 2.0E-14 | n.d. | | 0.183 | 0.003 | 6.78E-08 | 3.E-10 | -108 | 3 |
| | #7 | 1.4E-14 | 2.E-15 | 9.6 | 0.6 | 0.090 | 0.015 | 3.105E-12 | 2.6E-14 | 15.1 | 1.2 | 0.186 | 0.002 | 5.57E-08 | 3.E-10 | -104 | 3 |
| | #8 | 1.2E-14 | 2.E-15 | 10.6 | 0.7 | 0.059 | 0.010 | 9.566E-13 | 9.7E-15 | 1.0 | 4.0 | 0.175 | 0.005 | 1.62E-08 | 2.E-10 | -66 | 9 |
| | #9 | 2.3E-14 | 2.E-15 | 10.2 | 0.4 | 0.054 | 0.005 | 8.174E-13 | 1.4E-14 | 2.7 | 4.7 | 0.179 | 0.006 | 1.03E-08 | 2.E-10 | -64 | 16 |
| | #10 | 2.6E-14 | 2.E-15 | 9.7 | 0.4 | 0.035 | 0.006 | 5.335E-13 | 1.0E-14 | 8.2 | 7.2 | 0.193 | 0.012 | 6.83E-09 | 2.E-10 | -48 | 28 |
| | Total | 1.6E-13 | 2.E-14 | 10.7 | 1.1 | 0.047 | 0.007 | 3.838E-11 | 1.6E-13 | 8.7 | 0.2 | 0.186 | 0.001 | 5.10E-07 | 1.E-09 | -89 | 7 |
| H-N3 (2.35 mg) | #1 | 2.7E-15 | 6.E-16 | 14.2 | 1.5 | 0.073 | 0.020 | 1.587E-13 | 2.4E-15 | 121.2 | 11.4 | 0.189 | 0.009 | 3.60E-08 | 1.E-10 | 25 | 3 |
| | #2 | 3.2E-15 | 1.E-15 | 13.5 | 1.3 | 0.068 | 0.022 | 3.953E-13 | 8.0E-15 | n.d. | | 0.196 | 0.007 | 1.43E-08 | 9.E-11 | 9 | 6 |
| | #3 | 6.0E-15 | 6.E-16 | 13.0 | 0.7 | n.d. | | 1.145E-12 | 9.2E-15 | n.d. | | 0.192 | 0.002 | 5.20E-09 | 8.E-11 | 18 | 14 |
| | #4 | 3.0E-15 | 9.E-16 | 11.0 | 1.1 | 0.210 | 0.042 | 3.824E-13 | 4.6E-15 | 4.9 | 4.5 | 0.180 | 0.005 | 2.29E-08 | 1.E-10 | -9 | 48 |
| | #5 | 3.4E-15 | 6.E-16 | 9.9 | 0.8 | 0.026 | 0.015 | 2.054E-13 | 3.6E-15 | n.d. | | 0.192 | 0.009 | 8.86E-10 | 8.E-11 | -117 | 138 |
| | #6 | 4.6E-15 | 1.E-15 | 8.5 | 0.7 | 0.060 | 0.013 | 3.213E-13 | 4.2E-15 | 183.6 | 6.0 | 0.180 | 0.005 | 9.28E-10 | 8.E-11 | 25 | 72 |
| | #7 | 1.5E-14 | 1.E-15 | 9.2 | 0.4 | 0.035 | 0.006 | 2.475E-12 | 1.9E-14 | n.d. | | 0.188 | 0.002 | 1.77E-08 | 1.E-10 | -96 | 5 |
| | #8 | 2.5E-14 | 1.E-15 | 10.1 | 0.3 | 0.034 | 0.003 | 6.185E-12 | 3.9E-14 | 0.6 | 0.3 | 0.191 | 0.001 | 5.66E-08 | 2.E-10 | -111 | 1 |
| | #9 | 4.16E-14 | 9.E-16 | 9.5 | 0.2 | 0.031 | 0.007 | 1.577E-11 | 9.4E-14 | 0.9 | 0.1 | 0.187 | 0.001 | 1.53E-07 | 5.E-10 | -113 | 3 |
| | #10 | 2.2E-14 | 1.E-15 | 10.3 | 0.4 | 0.030 | 0.004 | 6.814E-12 | 4.3E-14 | 0.9 | 0.3 | 0.189 | 0.002 | 7.87E-08 | 3.E-10 | -108 | 3 |
| | #11 | 2.2E-14 | 1.E-15 | 8.7 | 0.3 | 0.025 | 0.006 | 1.139E-12 | 1.0E-14 | 3.0 | 1.5 | 0.188 | 0.002 | 2.43E-08 | 1.E-10 | -101 | 4 |
| | #12 | 1.47E-14 | 9.E-16 | 8.9 | 0.4 | 0.045 | 0.004 | < blk | | n.d. | | n.d. | | 4.67E-10 | 8.E-11 | -166 | 122 |
| | Total | 1.57E-13 | 9.E-15 | 11.1 | 0.7 | 0.040 | 0.005 | 3.499E-11 | 1.1E-13 | 3.0 | 0.1 | 0.188 | 0.001 | 4.11E-07 | 7.E-10 | -86 | 1 |

# Table 3

| Sample | Extraction step | $^{4}$He (mol/g) | ± | $^{3}$He/$^{4}$He | ± | $^{22}$Ne (mol/g) | ± | $^{20}$Ne/$^{22}$Ne | ± | $^{21}$Ne/$^{22}$Ne | ± | $^{36}$Ar (mol/g | ± | $^{40}$Ar/$^{36}$Ar | ± | $^{36}$Ar/$^{38}$Ar | ± |
|---|---|---|---|---|---|---|---|---|---|---|---|---|---|---|---|---|---|
| H-Z1 (0.5 mg) | total | 2.5E-10 | 4.9E-11 | 2.1E-04 | 7.E-05 | < blk | | n.d. | | | | 1.8E-11 | 8.E-12 | n.d. | | 0.2 | 0.1 |
| H-Z2 (2.9 mg) | total | 2.9E-10 | 1E-11 | 1.39E-04 | 1.E-05 | 3.3E-14 | 3.E-15 | 11 | 1 | 0.05 | 0.01 | 2.2E-11 | 1.E-12 | n.d. | | 0.20 | 0.01 |
| H-Z3 (2.9 mg) | total | 6.8E-10 | 3.5E-11 | 1.65E-04 | 9.E-06 | 4.3E-14 | 3.E-15 | 11.0 | 0.9 | 0.032 | 0.004 | 3.27E-11 | 8.E-13 | 2 | 2 | 0.19 | 0.01 |
| H-Z4 (1.0 mg) | total | 5.0E-10 | 3.5E-11 | 1.7E-04 | 2.E-05 | 3.3E-14 | 9.E-15 | 12 | 4 | 0.04 | 0.02 | 3.0E-11 | 2.E-12 | 0.1 | 7.6 | 0.19 | 0.02 |
| H-Z5 (2.8 mg) | 1000°C | 1.55E-11 | 2.9E-13 | | | < blk | | | | | | 1.10E-13 | 4.E-13 | | | | |
| | 1800°C | 6.2E-10 | 3.2E-11 | | | 3.6E-14 | 3.E-15 | | | | | 3.47E-11 | 9.E-13 | | | | |
| | total | 6.4E-10 | 3.2E-11 | 1.65E-04 | 9.E-06 | 3.6E-14 | 3.E-15 | 11 | 1 | 0.028 | 0.005 | 3.48E-11 | 1.E-12 | n.d. | n.d. | 0.19 | 0.01 |

**Table 4**

| **Hypatia** **(this study)** | Ureilites | Iron meteorites | Acapulcoites |
|---|---|---|---|
| **Size: cm** | mm | cm | µm |
| **Diamonds** | ✓ | ✓ | ✖ |
| **Q noble gases** | ✓ | ✓ | ✓ |
| **Excess $^{129}$Xe** | ✖ | ✓ | ✓ |
| **$\delta^{13}$C ≈ -3.4 ‰** | ✓ | – | ✖ |
| **Main $\delta^{15}$N ≈ -110 ‰** | ✓ | – | – |

# A comprehensive study of noble gases and nitrogen in "Hypatia", a diamond-rich pebble from SW Egypt

# Supplementary information

G. Avice[a*], M. M. M. Meier[b, a], B. Marty[a], R. Wieler[b], J. D. Kramers[c], F. Langenhorst[d], P. Cartigny[e], C. Maden[b], L. Zimmermann[a], M. A. G. Andreoli[f]

*corresponding author. E-mail: gavice@crpg.cnrs-nancy.fr

[a]CRPG-CNRS, Université de Lorraine, 15 rue Notre-Dame des Pauvres, BP 20, 54501 Vandoeuvre-lès-Nancy Cedex, France.

[b]Department of Earth Sciences, ETH Zürich, Clausiusstrasse 25, CH-8092 Zürich, Switzerland.

[c]Department of Geology, University of Johannesburg, Auckland Park 2006, South Africa.

[d]Institut für Geowissenschaften, Friedrich-Schiller-Universität Jena, Carl-Zeiss-Promenade 10, D-07745 Jena, Germany.

[e]Laboratoire de Géochimie des Isotopes Stables de l'Institut de Physique du Globe de Paris, UMR 7154, Université Paris Denis-Diderot, 1 rue Jussieu, 75005 Paris, France.

[f]School of Geosciences, University of the Witwatersrand, PO Box 3, Wits 2050, South Africa.

**Methods: X-ray diffraction and Transmission Electron Microscopy (TEM)**

Mineralogical characterization of Hypatia fragments was done by X-ray powder diffraction and TEM, using a Seifert-FPM XRD7 diffractometer (Cu-$K_\alpha$ radiation) and a 200 kV ZEISS LEO922 TEM at the University of Jena (Germany). For comparison, we applied the same techniques on impact diamonds from the Popigai crater (Koeberl et al., 1997), where impact diamonds were first discovered in terrestrial impact rocks (Masaitis et al., 1972). Samples were crushed in liquid nitrogen to obtain a powder for the X-ray diffraction experiment. The powder X-ray diffraction experiment on Hypatia was conducted for three days in order to obtain a pattern with a signal to noise ratio higher than 50 for the 111 reflection of diamond. Small fractions of the crushed samples were loaded on perforated carbon grids for TEM observations. Conventional bright-field/dark-field and high-resolution TEM imaging were used to characterize the micro-structure of Hypatia, which in turn provides clues to its deformation and transformation effects (Langenhorst and Deutsch, 2012; Langenhorst et al., 1999).

## Supplementary figures

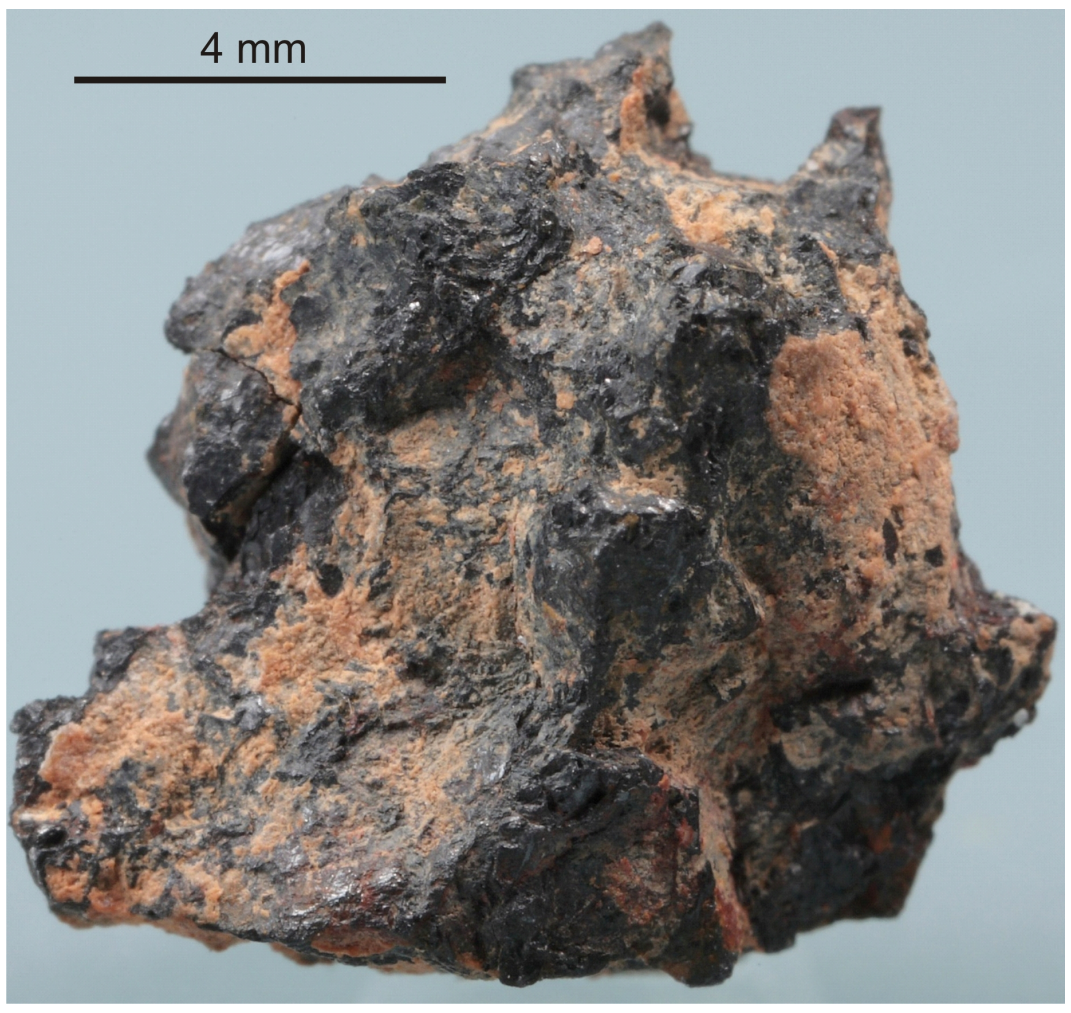

**Supplementary Figure S1: Photography of Hypatia** (Kramers et al., 2013)

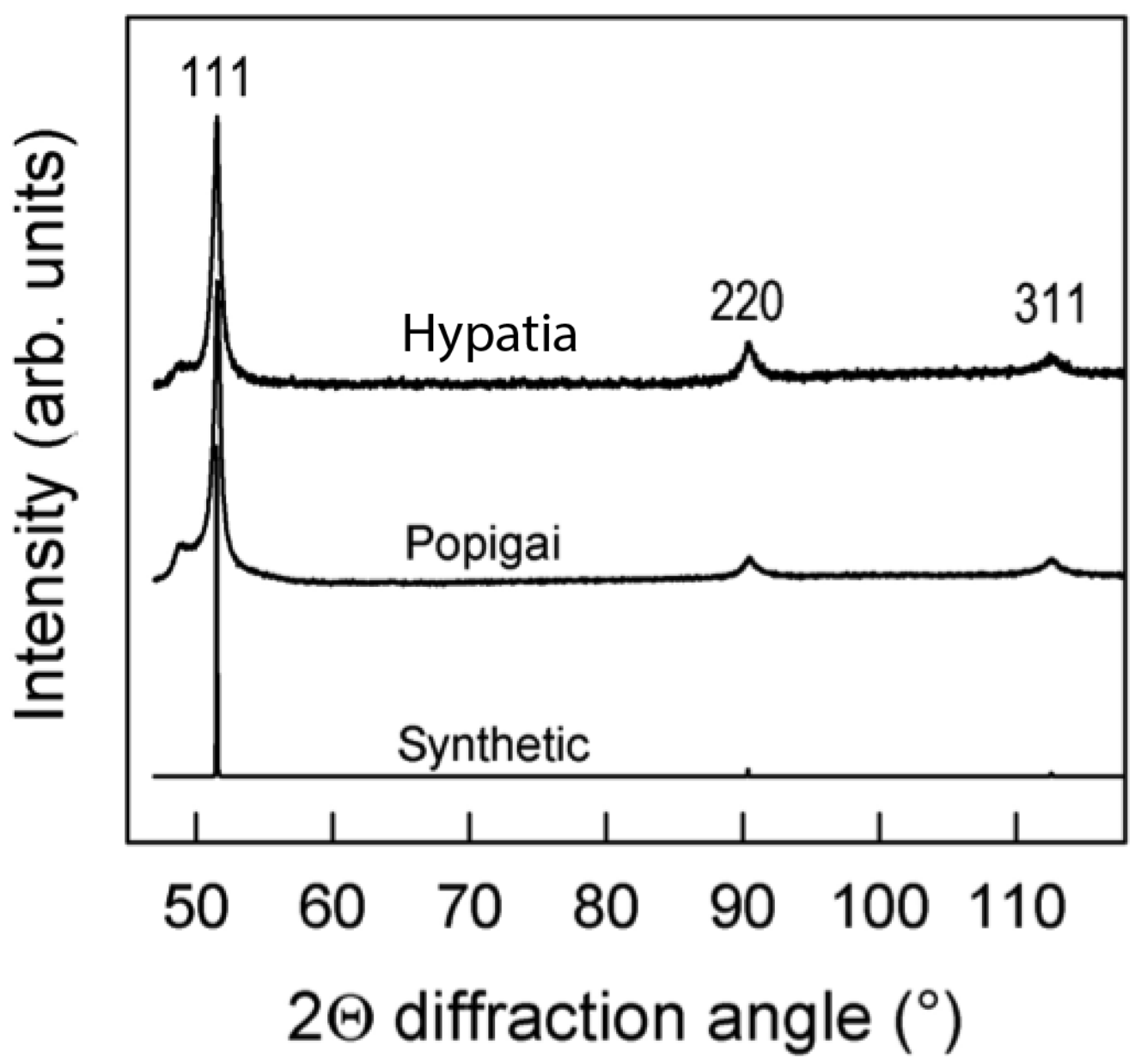

**Supplementary Figure S2: X-ray diffraction pattern of Hypatia in comparison to synthetic diamonds and impact diamonds from the Popigai structure.**

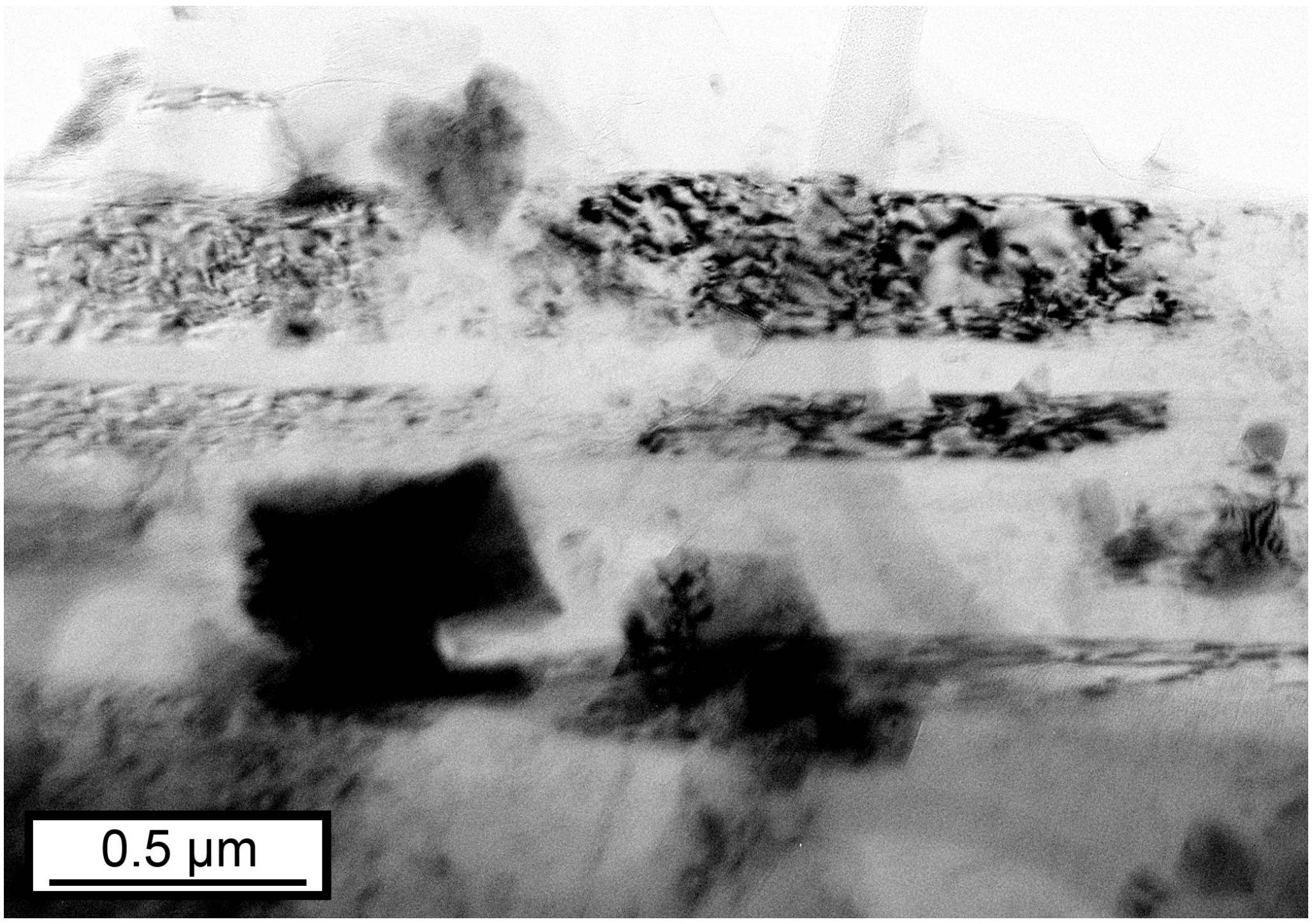

**Supplementary Figure S3: Bright-field TEM image of multiple deformation bands in Hyptia.**

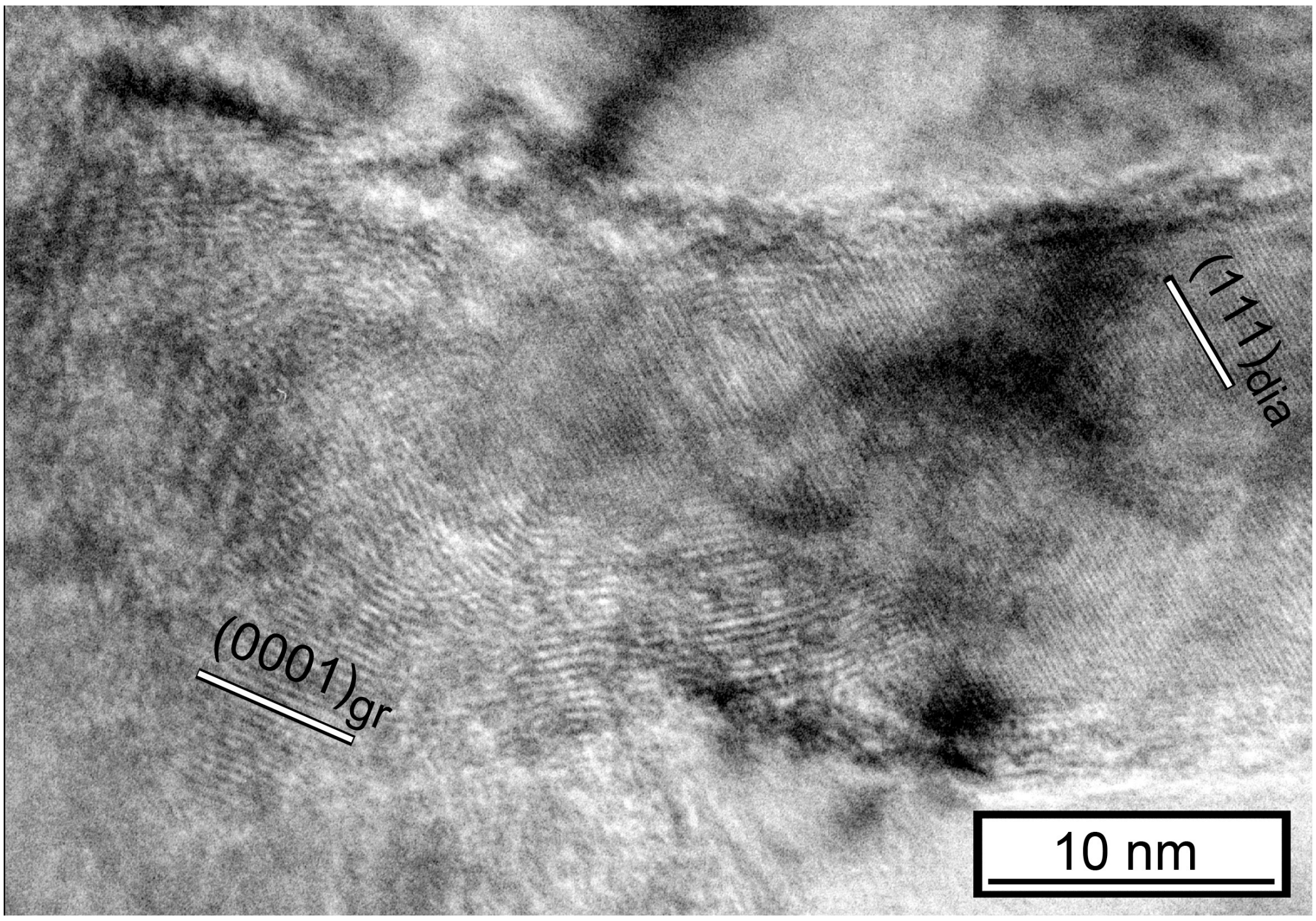

**Supplementary Figure S4: High resolution TEM image showing the onion-shaped graphite $(0001)_{gr}$ in the surface of diamond $(111)_{dia}$. The absence of orientation relationship suggests that graphite is here a product of retrograde annealing of diamond.**

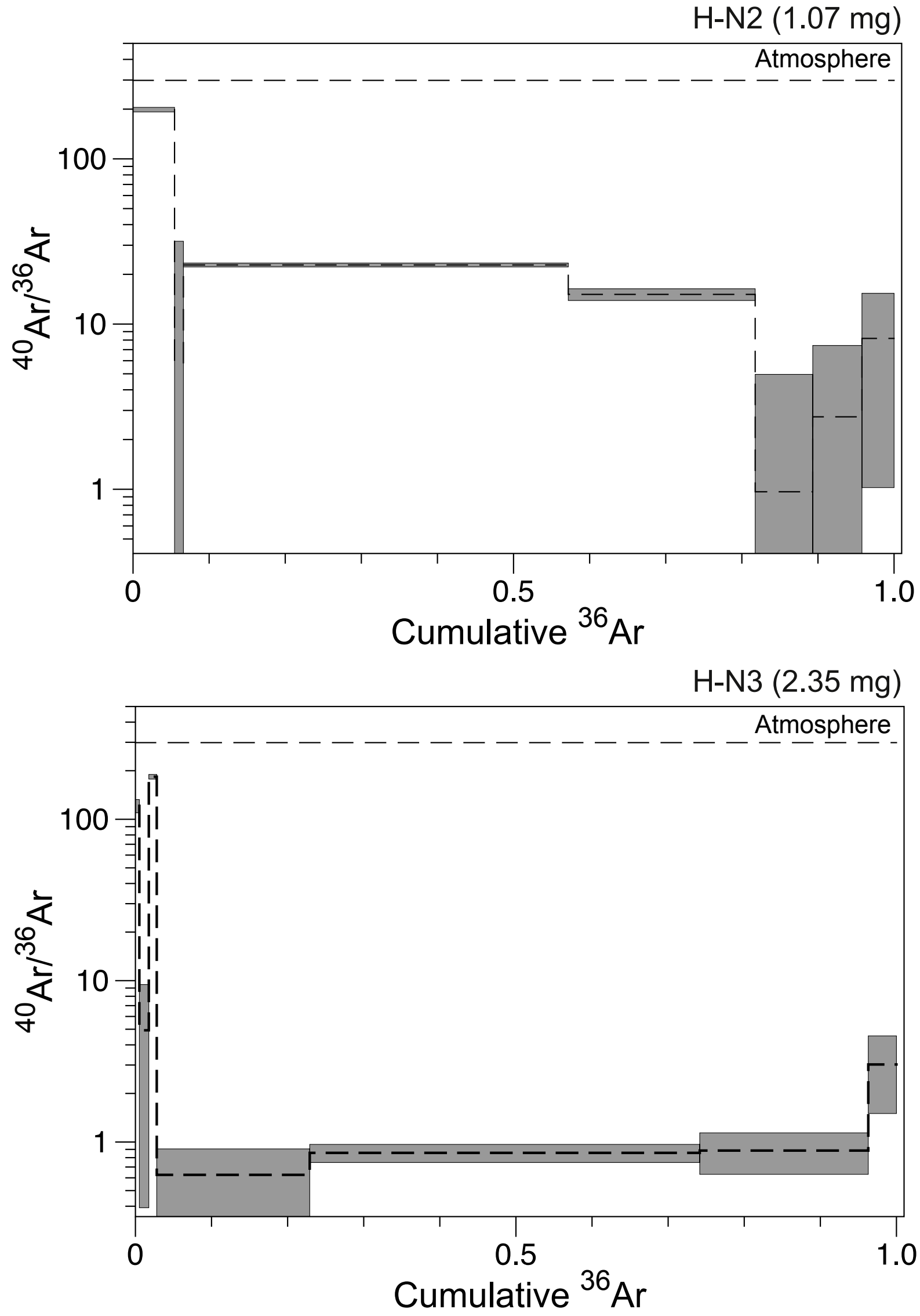

**Supplementary Figure S5: Evolution of the $^{40}Ar/^{36}Ar$ ratio during step-heating of the H-N2 and H-N3 samples. Ranges are at 1σ.**

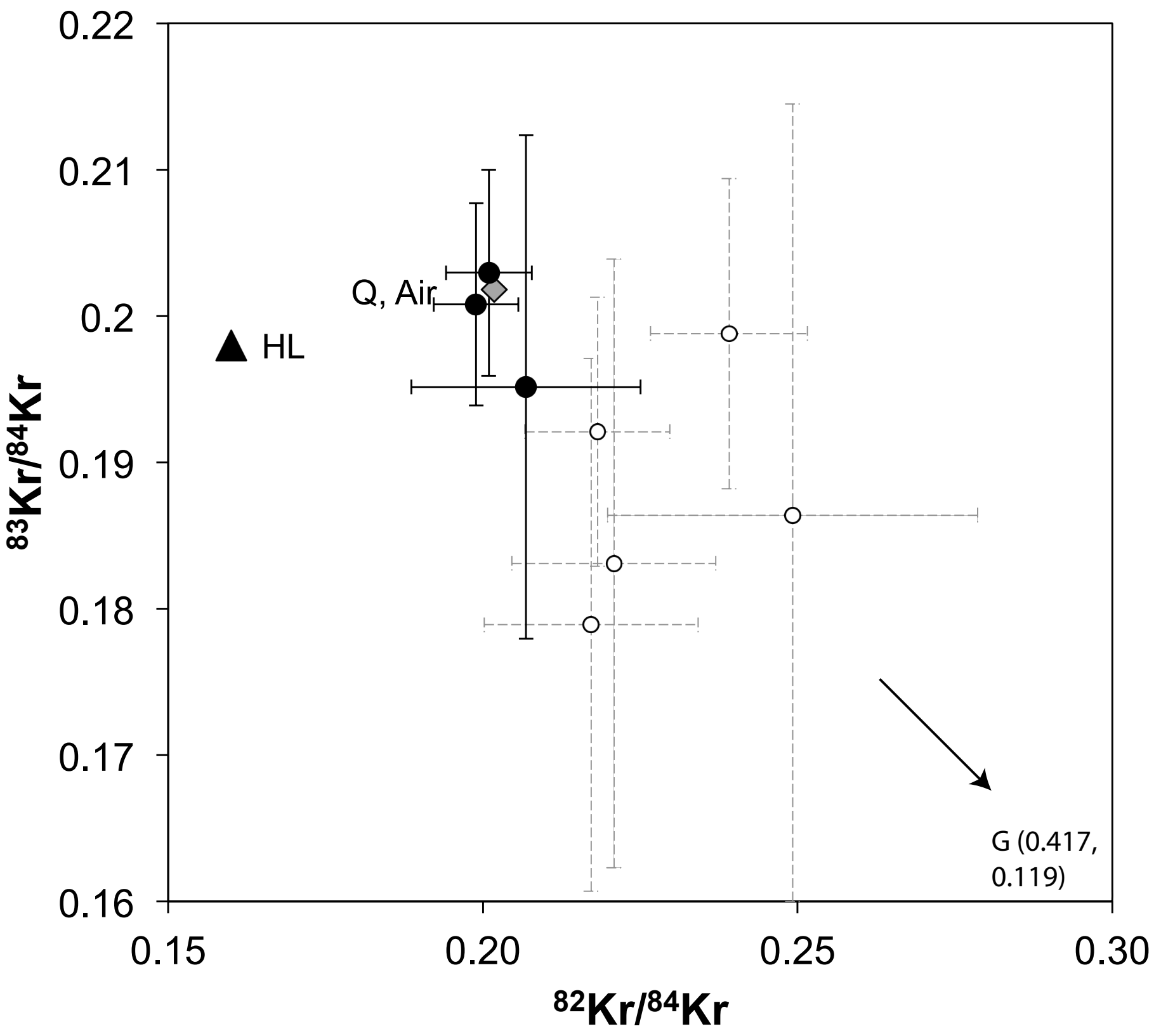

**Supplementary Figure S6: Three-isotope diagram of krypton. Heating steps of H-N4 are shown with black dots and Q, Air and HL isotopic compositions are from** (Ott, 2014). **Withe dots with dotted error bars are extreme data taken from** (Kramers et al., 2013). **Error bars are 2σ.**

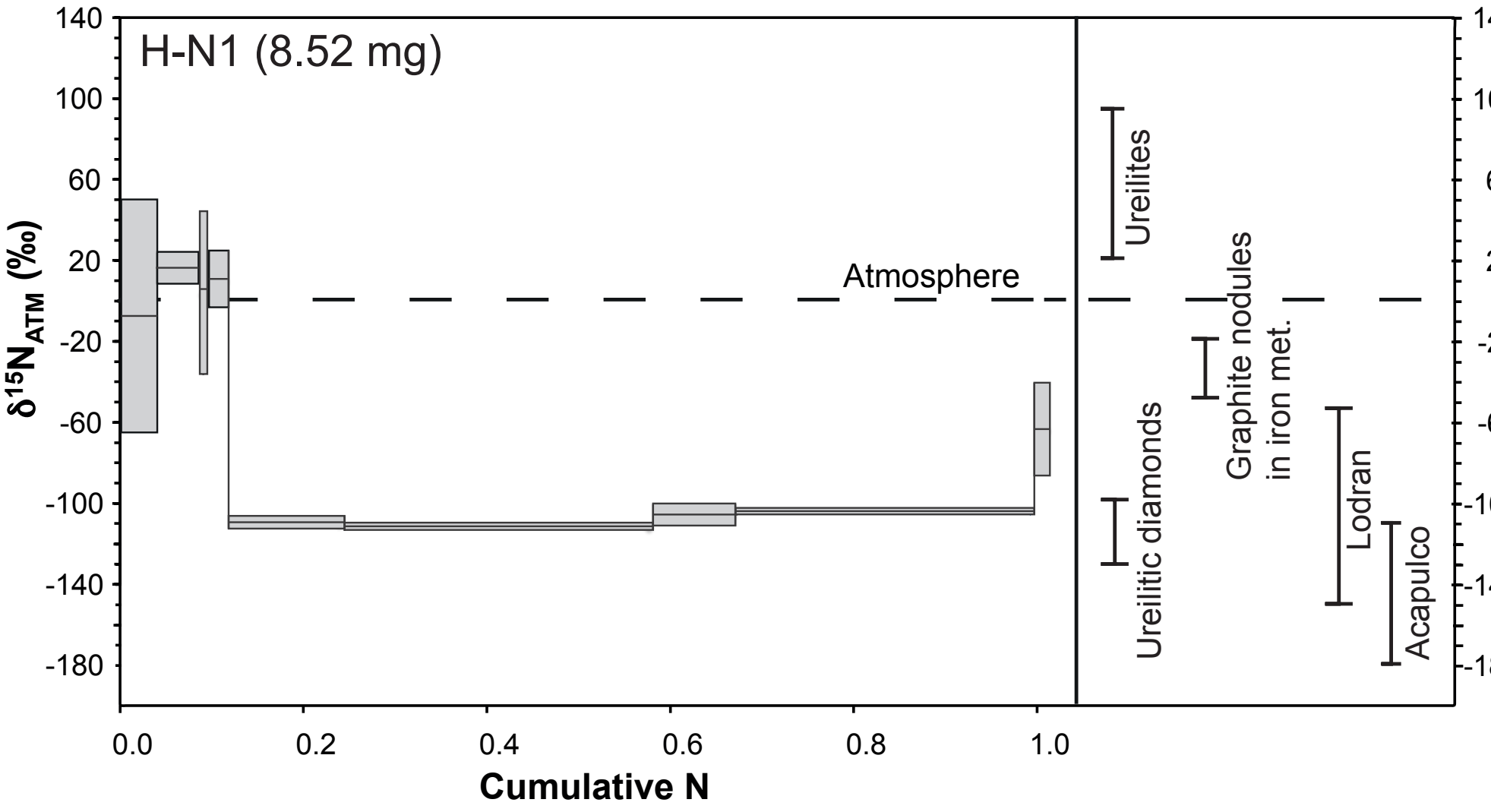

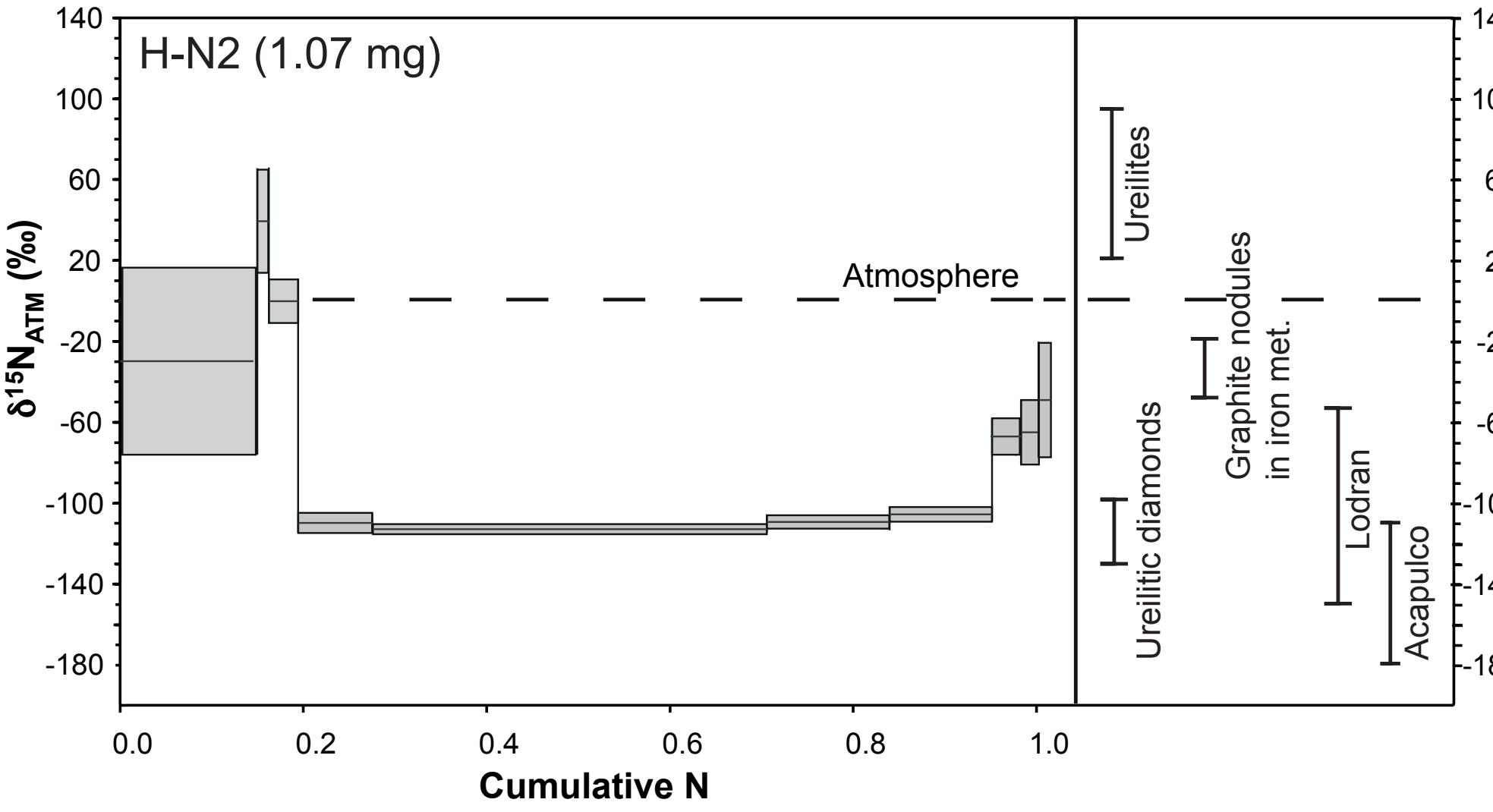

**Supplementary Figure S7: Ranges of isotopic compositions of nitrogen released during step-heating of the H-N1 (a) and H-N2 (b) samples. Ranges for individual steps correspond to errors at 1σ. Ranges of isotopic composition of N in C-rich meteoritic materials are shown for comparison. See main text for references.**

**Supplementary Table S1: Data of bulk N and C isotope measurements conducted in IPG-Paris. Errors at 2σ.**

| Sample | Weight (mg) | $\delta^{15}N$ (‰) | ± | N (ppm) | ± | $\delta^{13}C$ | ± | % C |
|---|---|---|---|---|---|---|---|---|
| H-P1 | 1.57 | -99.6 | 1 | 17.1 | 1.71 | -3.54 | 0.14 | 93 |
| H-P2 | 1.61 | -99.9 | 1 | 19.3 | 1.93 | -3.33 | 0.14 | 95 |

# Supplementary Table S2: Data of Kr and Xe measured in Hypatia sample H-N4. Errors at 2σ.

**H-N4 (8.4 mg)**

| T (°C) | $^{132}$Xe (mol/g) | ± | $^{124}$Xe/$^{132}$Xe | ± | $^{126}$Xe/$^{132}$Xe | ± | $^{128}$Xe/$^{132}$Xe | ± | $^{129}$Xe/$^{132}$Xe | ± | $^{130}$Xe/$^{132}$Xe | ± | $^{131}$Xe/$^{132}$Xe | ± | $^{134}$Xe/$^{132}$Xe | ± | $^{136}$Xe/$^{132}$Xe | ± |
|---|---|---|---|---|---|---|---|---|---|---|---|---|---|---|---|---|---|---|
| 400 | 8.05E-16 | 3.E-17 | 0.0024 | 0.0008 | 0.0035 | 0.0009 | 0.076 | 0.007 | 1.00 | 0.04 | 0.143 | 0.010 | 0.790 | 0.035 | 0.396 | 0.017 | 0.317 | 0.021 |
| 850 | 9.67E-16 | 3.E-17 | 0.0040 | 0.0008 | 0.0032 | 0.0008 | 0.071 | 0.005 | 1.00 | 0.03 | 0.157 | 0.011 | 0.778 | 0.031 | 0.375 | 0.013 | 0.327 | 0.014 |
| 1400 | 1.24E-14 | 2.E-16 | 0.0047 | 0.0003 | 0.0039 | 0.0003 | 0.083 | 0.003 | 1.19 | 0.04 | 0.162 | 0.006 | 0.819 | 0.027 | 0.382 | 0.013 | 0.314 | 0.011 |
| 1800 | 3.35E-15 | 1.E-16 | 0.0042 | 0.0006 | 0.0040 | 0.0005 | 0.083 | 0.005 | 1.18 | 0.05 | 0.160 | 0.008 | 0.805 | 0.034 | 0.373 | 0.016 | 0.315 | 0.014 |
| 2200 | 2.22E-14 | 3.E-16 | 0.0045 | 0.0003 | 0.0040 | 0.0003 | 0.082 | 0.002 | 1.19 | 0.01 | 0.161 | 0.003 | 0.820 | 0.008 | 0.380 | 0.005 | 0.315 | 0.005 |
| 2200 bis | 2.96E-15 | 7.E-17 | 0.0047 | 0.0007 | 0.0035 | 0.0006 | 0.083 | 0.005 | 1.15 | 0.04 | 0.162 | 0.007 | 0.809 | 0.033 | 0.384 | 0.015 | 0.321 | 0.013 |
| Total | 4.27E-14 | 4.E-16 | 0.0045 | 0.0003 | 0.0039 | 0.0003 | 0.082 | 0.003 | 1.18 | 0.03 | 0.161 | 0.005 | 0.816 | 0.021 | 0.380 | 0.010 | 0.315 | 0.009 |

| T (°C) | $^{84}$Kr (mol/g) | ± | $^{80}$Kr/$^{84}$Kr | ± | $^{82}$Kr/$^{84}$Kr | ± | $^{83}$Kr/$^{84}$Kr | ± | $^{86}$Kr/$^{84}$Kr | ± |
|---|---|---|---|---|---|---|---|---|---|---|
| 400 | 8.6E-16 | 1.E-16 | 0.048 | 0.010 | 0.207 | 0.018 | 0.195 | 0.017 | 0.31 | 0.03 |
| 850 | 7.2E-16 | 2.E-16 | 0.043 | 0.041 | 0.204 | 0.135 | 0.200 | 0.117 | 0.30 | 0.18 |
| 1400 | 5.10E-15 | 3.E-16 | 0.037 | 0.004 | 0.201 | 0.007 | 0.203 | 0.007 | 0.31 | 0.01 |
| 1800 | 4.61E-15 | 1.E-15 | 0.030 | 0.053 | 0.194 | 0.266 | 0.202 | 0.260 | 0.31 | 0.40 |
| 2200 | 3.7E-14 | 2.E-15 | 0.036 | 0.004 | 0.199 | 0.007 | 0.201 | 0.007 | 0.31 | 0.01 |
| 2200 bis | 7.72E-15 | 3.E-15 | 0.037 | 0.116 | 0.184 | 0.579 | 0.203 | 0.572 | 0.31 | 0.87 |
| Total | 5.6E-14 | 4.E-15 | 0.036 | 0.018 | 0.198 | 0.087 | 0.201 | 0.085 | 0.31 | 0.13 |